\documentclass[12pt, a4paper]{article}
\usepackage{color}

\usepackage{latexsym}
\usepackage{amsmath, amscd}
\usepackage{multirow}
\usepackage{amssymb}
\usepackage{graphicx}
\usepackage{euscript}
\usepackage{amsfonts}
\usepackage{verbatim}
\usepackage{epsfig}
\usepackage{hyperref}
\usepackage[margin=2cm,bottom=2.0cm]{geometry}

\def\be{\begin{eqnarray}}
\def\ee{\end{eqnarray}}
\def\0{\nonumber}
\def\d{\partial}

\providecommand{\IIm}{\textnormal{Im}}

\providecommand{\ch}{\textnormal{ch}}
\providecommand{\PD}{\textnormal{PD}}

\providecommand{\sq}{\textnormal{sq}}
\providecommand{\PP}{\mathbb{P}}

\def\0{\nonumber}

\def\cO{\mathcal{O}}
\def\cL{\mathcal{L}}

\begin{document}

\begin{titlepage}
\titlepage
\begin{flushright}{MAD-TH-10-09 \\
LMU-ASC 100/10}
\end{flushright}
\vskip 2.5cm
\centerline{ \bf \huge On Flux Quantization in F-Theory}
\vskip 2.3cm

\begin{center}
{\bf \large Andr\'es Collinucci}
\vskip 0.3cm
\em 
Arnold Sommerfeld Center for Theoretical Physics\\
Ludwig-Maximilians-Universit\"at M\"unchen\\
Theresienstrasse 37, 80333 M\"unchen, Germany

\vskip 1.2cm

{\bf \large Raffaele Savelli}
\vskip 0.3cm

Department of Physics, University of Wisconsin-Madison\\
1150 University Avenue Madison, WI 53706-1390, USA

\vskip 2.5cm

\large \bf Abstract
\end{center}

\normalsize We study the problem of four-form flux quantization in F-theory compactifications. We prove that for smooth, elliptically fibered Calabi-Yau fourfolds with a Weierstrass representation, the flux is always integrally quantized. This implies that any possible half-integral quantization effects must come from 7-branes, i.e. from singularities of the fourfold.

We subsequently analyze the quantization rule on explicit fourfolds with $Sp(N)$ singularities, and connect our findings via Sen's limit to IIB string theory. Via direct computations we find that the four-form is half-integrally quantized whenever the corresponding 7-brane stacks wrap non-spin complex surfaces, in accordance with the perturbative Freed-Witten anomaly. Our calculations on the fourfolds are done via toric techniques, whereas in IIB we rely on Sen's tachyon condensation picture to treat bound states of branes.
Finally, we give general formulae for the curvature- and flux-induced D3 tadpoles for general fourfolds with $Sp(N)$ singularities.

\vskip2cm

\vskip1.5\baselineskip

\vfill
\vskip 2.mm
\end{titlepage}

\tableofcontents

\section{Introduction}

F-theory \cite{Vafa:1996xn}, which describes non-perturbative IIB backgrounds geometrically, can be related to the more physical picture of M-theory. The precise relation in terms of fiber-wise dualities (explained in \cite{Denef:2008wq, Grimm:2010ks, Weigand:2010wm}) also tells us that both bulk three-form and 7-brane worldvolume fluxes are encoded in the M-theory four-form flux $G_4$.

The framework of local F-theory model building \cite{Beasley:2008dc, Beasley:2008kw, Donagi:2008ca, Donagi:2008kj, Donagi:2009ra} has brought on numerous important insights that were necessary from a phenomenological point of view.
On the other hand, it has become apparent that understanding compact models is crucial to make the picture consistent and to ensure that the desirable ingredients of the local perspective truly are present. For instance, admitting a `decoupling limit'  \cite{Cordova:2009fg}, or computing a non-perturbative superpotential \`a la \cite{Witten:1996bn} for moduli stabilization \cite{Blumenhagen:2010ja, Cvetic:2010ky, Cvetic:2010rq}, all require control over compact manifolds. Since then, many interesting compact Calabi-Yau \emph{fourfolds} have been explicitly constructed \cite{Collinucci:2009uh, Blumenhagen:2009yv, Grimm:2009yu, Chen:2010ts, Cvetic:2010rq}. 

Nevertheless, a systematic treatment of $G_4$ fluxes is still elusive. One aspect that has not been explored, is its quantization. 
In his seminal work \cite{Witten:1996md}, Witten derived the general rule for flux quantization in M-theory:
\be
G_4-\frac{\lambda}{2}\,\in\,H^4(M_{11},\mathbb{Z})\,,
\ee
where $\lambda$ is one half the first Pontryagin class of the eleven-dimensional spacetime. This rule implies that, depending on the topology of spacetime, the quantization of $G_4$ may be shifted by a half-integer. If that is the case, then $G_4$ cannot be set to zero. 

In F-theory compactifications to four dimensions, this can have several consequences: Firstly, in models with intersecting 7-branes, such an obligatory half-integral flux may correspond in IIB to a worldvolume flux, which in turn may affect the spectrum of 4d chiral fermions. Secondly, for smooth elliptically fibered CY fourfolds, a flux containing a $\lambda/2$ component maps to a IIB flux that breaks 4d Poincar\'e invariance \cite{Dasgupta:1999ss, Denef:2008wq}. It is therefore crucial to gain systematic understanding of when this comes about.

One expects that, in general, whenever the CY fourfold requires that the quantization of $G_4$ flux be shifted, this will map to some shifted worldvolume flux on a non-spin D7-brane in IIB string theory.  However, such a statement is ill-defined since the clear distinction between IIB worldvolume fluxes and bulk fluxes becomes blurred in F-theory. As explained in great detail in \cite{Denef:2008wq}, mapping a $G_4$ flux to IIB is only well defined up to monodromy effects. More precisely, an M-theory flux that maps to a 7-brane worldvolume will pick up a bulk $(H_3, F_3)$ piece upon moving said 7-brane around in its open string moduli space. 
Secondly, contrary to the intuitive belief that $G_4$ fluxes corresponding to brane fluxes are always sharply localized around the discriminant locus, worldvolume fluxes on Abelian stacks can be completely delocalized once $g_s$ is turned on (see \cite{Braun:2011zm} for an example of such fluxes).
Hence, the perturbative meaning of half-integrally quantized $G_4$ flux requires more careful analysis than one might expect.

In this paper, we first analyze flux quantization in the case of smooth fourfolds and then in the case with non-Abelian singularities in the $C$-series (i.e. $Sp(N)$). In order to interpret our results, we relate F-theory to IIB string theory via Sen's limit, and then treat all D7-branes in terms of D9/anti-D9 tachyon condensates by expanding upon the framework developed in \cite{Collinucci:2008pf}. 

We find that, for smooth elliptic fibrations, \emph{$G_4$ is always integrally quantized}. This implies that half-integral quantization effects can only come from (resolved) singularities in the CY fourfold. Put differently, our results show that any possible half-integrally quantized $G_4$ fluxes must be localized on stacks of 7-branes.

In order to study these effects, we construct fourfolds with $C_n$-type singularities, on which we perform blow-ups explicitly with the help of the toric package in SAGE \cite{sage}. This allows us to compute $\lambda$, and hence deduce the flux quantization rule.

By using the tachyon condensation/K-theory picture, we compare the quantization of $G_4$ to the quantization of worldvolume flux on the corresponding D7-branes with $Sp(N)$ gauge groups. We find that $G_4$ is indeed half-integrally quantized whenever the D7-branes wrap non-spin complex surfaces, as expected from the Freed-Witten anomaly \cite{Freed:1999vc}.

This work is a first step in analyzing how the perturbative Freed-Witten anomaly generalizes to F-theory. Other ramifications include studying flux quantization on M5-branes, self-duality conditions on IIB bulk fluxes, and quantization conditions on IIA fluxes. 

\vskip3mm

This paper is organized as follows: In section \ref{settingproblem} we introduce the issue of flux quantization in M-theory, thereby setting up the problem of interest. In section \ref{casoliscio} we prove that, in smooth elliptic fibrations with global $E_8$ Weierstrass representations\footnote{We also prove this for the cases with global $E_7$ and $E_6$ representations.}   the flux is always integrally quantized. In section \ref{necessaryB}, we analyze the perturbative IIB situation linked via Sen's limit to F-theory. We specifically study two possible sources for half-integral quantization in IIB when non-Abelian gauge groups are absent: O7-planes on non-spin complex surfaces, and O3-planes. We show that the former do not lead to half-integral quantization, and conjecture that the latter do not either.

In section \ref{SingularCase} we move to the case of singular CY fourfolds. We explicitly construct a fourfold with an $Sp(1)$ singularity, and compute the resulting $\lambda$ and Euler characteristic. We then compare the results on flux quantization with the perturbative IIB counterparts by using the tachyon condensation picture, where D7's are treated as D9/anti-D9 pairs. We find, that whenever a blow-up is performed, the $\lambda$ and the Euler characteristic of the fourfold always change such that the induced D3-charge
\begin{equation}
Q_{D3} = \frac{\chi(Z_4)}{24} - \frac{1}{2}\,\int_{Z_4} \left(\frac{\lambda}{2} \right)^2
\end{equation}
remains constant. We explain this physically by showing that the process of creating a singularity (and blowing it up) corresponds in IIB to the process where a single D7-brane, carrying the total D7-charge, decays into a non-Abelian stack plus a `remainder' D7-brane. In section \ref{boundstates} we show how such processes of brane recombination/separation take place in terms of the tachyon condensation picture.

In the remainder of section \ref{SingularCase}, we generalize our results to fibrations over a general K\"ahler base threefold. By using the IIB tachyon condensation picture, we generalize our results to $Sp(N)$ singularities, thereby giving a general formula for the curvature-induced and the flux-induced D3 tadpoles. Our general formulae are checked in our specific working example by calculations performed on SAGE for $Sp(N)$ with $N=1, \ldots, 4$.
We defer the details of the toric calculations to the appendices.

In section \ref{summaryoutlook}, we briefly summarize our results and main formulae, and conclude with an outlook.

\section{Setting up the problem}\label{settingproblem}

The Freed-Witten anomaly has a double impact on the string background: On the one hand it imposes a necessary and sufficient condition to anomaly cancellation in terms of torsion characteristic classes. On the other hand, it gives the right quantization condition to be imposed on fluxes in order to actually get rid of the anomaly, once the background fulfills the former condition. In this paper, we address the second aspect in the context of F-theory, by exploiting its duality with M-theory.

In M-theory a phenomenon similar to the global ambiguity of the string path integral measure arises, but now in the context of the quantum theory of membranes \cite{Witten:1996md}. Indeed, let the 11-dimensional target space of M-theory be the spin manifold $M_{11}$. The property of being spin implies that the first Pontryagin class $p_1(M_{11})$ is an even class\footnote{Notice that the converse of this statement is not true, because $p_1\;\textrm{mod}\,2=w_2^2$, which can very well be vanishing even though the manifold admits no spin structures (i.e. $w_2\neq0$).} in the fourth cohomology of $M_{11}$: Let 
\begin{equation} \label{lambdadef} \lambda(M_{11})=p_1(M_{11})/2\,. \end{equation} 

Now, an M2-brane has a 3-dimensional worldvolume and M-theory is just the theory of the embeddings of this manifold in $M_{11}$. Such a non-linear sigma model is taken to be supersymmetric but, unlike the case of strings, there is no chirality for the worldvolume spinors. Hence, while for the closed string path integral, the ambiguity of the Pfaffian of the world-sheet Dirac operator cancels between left and right movers \cite{Freed:1999vc}, here an ambiguity does survive and can be shown to be related to topological properties of the target as follows. Bringing the membrane worldvolume along a non trivial circle $C$, the Pfaffian of the Dirac operator comes back to itself up to the sign
\be
(-1)^{(\lambda\,,\,M2\times C)}\,,
\ee
where the round brackets in the exponent just mean evaluation of the class $\lambda$ on the 4-dimensional manifold $M2\times C$. The only possibility to compensate for such a global anomaly in order to render the partition function of the theory well-defined is to require an equal ambiguity for the holonomy of the $C_3$ potential over the M2 worldvolume that also enters the path integral measure. This is achieved by simply requiring a suitable quantization condition for the $G_4$ field strength\footnote{Square brackets indicating the cohomology classes are dropped, as well as factors of $2\pi$.}:
\be\label{WittenQuantization}
G_4-\frac{\lambda}{2}\,\in\,H^4(M_{11},\mathbb{Z})\,.
\ee 
Notice, in particular, that this anomaly can always be canceled, namely there are no global obstructions, but a suitable quantization condition for the G-flux must be chosen to get a consistent quantum theory. Moreover, the fact that closed strings never have this problem for their path integral implies that the H-flux, which in turn is responsible for the well-definiteness of the holonomy of the B-field over the world-sheet, must be always an integral class in the third cohomology of the string target $M_{10}$.

This quantization condition has consequences for the quantization of type IIA bulk fluxes, which we will not discuss here. However, it also has important consequences for the worldvolume flux quantization of D6-branes, as explored in \cite{Gukov:2001hf, Sparks:2003ck}.

Namely, suppose $M_{11}$ is a non-trivial $S^1$-fibration over $M_{10}$ giving rise to a so-called Taub-NUT geometry. Then, the codimension 3 loci of the base on which the $S^1$ fiber collapses are interpreted as D6-branes of type IIA string theory. The shift in the quantization of the M-theory $G_4$ flux directly induces a shift \`a la Freed-Witten \cite{Freed:1999vc, Minasian:1997mm} in the quantization condition of the gauge flux $F$ on the D6-branes. This phenomenon is naturally expected to happen because the open strings responsible for the possible half-integral quantization of $F$ lift to \emph{closed} M2-branes, since the $S^1$ is collapsing on the boundaries of the open strings (the D6-branes loci). Therefore the ambiguity in the path integral measure of such open strings gets directly related to the above mentioned one in the path integral measure of closed membranes leading to \eqref{WittenQuantization}. \\

The essence of this paper will be the search for the analogous effect on the gauge flux on the F-theory 7-branes. We will use the setting of elliptically fibered CY fourfolds for two reasons:  They provide a relevant setting for model building; and we can put the powerful machinery of algebraic geometry to work for us, which will greatly simplify the computations.

The phenomenon of half-integral quantization of $G_4$ should signal the presence of a Freed-Witten anomaly on the F-theory 7-brane that is cancelled by requiring the worldvolume gauge field to be a connection on a ``half''-line bundle.  The context is described by means of an $M_{11}$ elliptically fibered over $B_3$, and the $G_4$ flux inducing brane-type flux as in \cite{Denef:2008wq}. In other words, a spin$^c$ bundle is supposed to arise on a 7-brane wrapping a non-spin divisor $S_2\subset B_3$. Hence, the gauge flux $F$ cannot be set to zero because of its half-integral quantization.

Notice, on the other hand, that the IIB bulk fluxes do not undergo such a shift in their quantization in order to preserve the well-definiteness of the path integral measure of fundamental strings and their S-dual D1-branes.

In this paper, we will focus on the $\lambda$ class of the F-theory internal manifold, that decides the right quantization condition for $G_4$ according to \eqref{WittenQuantization}. Since we will consider F-theory on elliptically fibered Calabi-Yau fourfolds, the following simplification can be made: For a complex manifold $X$ there is a nice explicit expression for its first Pontryagin class in terms of its Chern classes:
\be
p_1(X)\,=\,c_1^2(X)-2\,c_2(X)\,.
\ee
This implies that, for a Calabi-Yau fourfold $Z_4$, $\lambda(Z_4)=-c_2(Z_4)$. Thus, the quantization condition of interest for $G_4$ becomes
\be\label{WittenQuantizationCY}
G_4+\frac{c_2(Z_4)}{2}\,\in\,H^4(Z_4,\mathbb{Z})\,,
\ee

We will address both smooth and singular $CY_4$'s, finding a number of new results about the Freed-Witten anomaly cancellation of the F-theory 7-branes. A general pattern for the form and for the oddness or evenness of $c_2(Z_4)$ will be also conjectured on the base of several concrete examples that will be constructed in detail and matched with models available in the literature.

\section{Smooth Calabi-Yau fourfolds}\label{casoliscio}

In this section, we will study in detail the second Chern class of a general, smooth, elliptically fibered Calabi-Yau fourfold $Z_4$, admitting a global Weierstrass representation. We will focus on Weierstrass models with $E_8$ fibrations, due to their connection with IIB string theory. However, our analysis can be repeated for $E_6$ and $E_7$ fibrations, whose weak coupling limits were constructed in \cite{Aluffi:2009tm}, yielding the same conclusion.
The strategy for deciding whether it is even or odd is to reduce the problem to a much easier one formulated on the base $B_3$ of the elliptic fibration, by using its Weierstrass form. 

The result of this section will be that \emph{the second Chern class of such Calabi-Yau manifolds is always even}. The physical consequences of this fact will be spelt out.

Smooth elliptically fibered Calabi-Yau fourfolds can be constructed in practice (even if sometimes it could be not sufficient) by taking for instance the base $B_3$ to be an \emph{almost Fano} threefold, i.e. its anti-canonical bundle is nef and `big'. Roughly, this means that the base has positive semi-definite curvature. If this requirement is not met, the anti-canonical bundle $K_{B_3}^{-1}$, and powers thereof may develop `base-loci' where all sections vanish simultaneously. This would lead to unwanted singularities of $Z_4$. Indeed, if $c_1(B_3)$ becomes negative on some 2-cycle, $K_{B_3}^{-|n|}$ will have no non-vanishing sections when restricted to that locus, and thus the polynomials appearing in the Kodaira representation of the Weierstrass form will vanish as well. This surely implies a singularity of the total space of the fibration.
 
Using standard formulae from algebraic geometry such as the \emph{adjunction formula}, as in \cite{Sethi:1996es}, it is easy to compute the total Chern class of a smooth Calabi-Yau fourfold described as a codimension one hypersurface by the Weierstrass polynomial:
\be\label{Kodairaparametr}
Y^2=X^3+fXZ^4+gZ^6\,.
\ee
Let $M_5$ be the ambient fivefold of $Z_4$, which is chosen to be a $W\PP^2_{2,3,1}$-bundle over $B_3$, whereby the fiber is parametrized by the weighted projective coordinates $X, Y$ and $Z$, of weights $2,3$ and $1$, respectively. 

Define $\mathcal{O}_{W\PP^2_{2,3,1}}(1)$ to be the `hyperplane' line bundle along the fiber, such that $Z$ is a section of it.  We will drop the subscript for convenience. This line bundle extends onto the whole projective bundle $M_5$ over $B_3$. Let $F\in H^2(M_5,\mathbb{Z})$ be the first Chern class of the line bundle\footnote{The pull-back of the projection map from $M_5$ to $B_3$ is always implicit in the notation.} $\mathcal{O}(1)\otimes K^{-1}(B_3)$, on $M_5$. 

Then, to make equation \eqref{Kodairaparametr} consistent, $X$, $Y$, $Z$, $f$, and $g$ must be sections of $\mathcal{O}(2)\otimes K^{-2}(B_3)$, $\mathcal{O}(3)\otimes K^{-3}(B_3)$, $\mathcal{O}(1)$, $K^{-4}(B_3)$ and $K^{-6}(B_3)$, respectively. The Weierstrass fibration admits a section, $Z=0$, whose Poincar\'e dual is, in this notation, $F-c_1(B_3)$. Now, recalling the general construction of an elliptically fibered Calabi-Yau, it is straightforward to deduce the following adjunction formula expressing the total Chern class of $Z_4$ in terms of $F$ and of the one of $B_3$:
\be\label{ChernclassZ4liscio}
c\,(Z_4)&=&\frac{c\,(B_3)\cdot(1+2F)\cdot(1+3F)\cdot(1+F-c_1(B_3))}{1+6F}\,.
\ee
The fact that $X, Y,$ and $Z$ parametrize the projective fiber means that they cannot vanish simultaneously in $M_5$. This is expressed easily by the following constraint:
\be\label{Constraint1}
F^2 \cdot (F-c_1(B_3))=0\,,
\ee
Since the Weierstrass equation \eqref{Kodairaparametr} defining $Z_4$ represents a divisor of class $6F$, on $Z_4$ the constraint \eqref{Constraint1} simply reduces to the condition $F^2=F\cdot c_1(B_3)$. Therefore, the second order in the expansion of \eqref{ChernclassZ4liscio}, namely the second Chern class of $Z_4$, will be:
\be\label{2ChernclassZ4liscio}
c_2(Z_4)&=&12F^2+c_2(B_3)-c_1^2(B_3)\,.
\ee

Before trying to understand, whether \eqref{2ChernclassZ4liscio} is even or odd, some comments on its structure are in order (see \cite{Marsano:2009ym} for similar observations). It turns out \cite{Dasgupta:1999ss, Denef:2008wq} that, in order to preserve Poincar\'e invariance in $\mathbb{R}^{1,3}$, the flux $G_4$ must have one and only one leg along the elliptic fiber. Now, since the second and the third term in \eqref{2ChernclassZ4liscio} are pull-backs from the base, they cannot have any legs along the fiber. The first term, on the other hand, can be written as $c_1^2(B_3)+c_1(B_3)\cdot(F-c_1(B_3))$, where the first term is again all along the base, while the second has two legs on the base and two on the fiber, whereby $F-c_1(B_3)$ is the Poincar\'e dual of the 0-section. Hence, the structure of $c_2(Z_4)$ is necessarily unacceptable  if one wants to keep space-time Poincar\'e invariance, and thus its possible oddness would induce a  Poincar\'e-breaking $G_4$ flux that cannot be put to zero because it is half-integrally quantized. 

As a general rule, on a smooth elliptic Calabi-Yau fourfold of strict SU(4) holonomy, any 4-cycle which is Poincar\'e dual to a Poincar\'e-preserving $G_4$ must be orthogonal to any intersection of two divisors. Indeed, apart from $c_2(B_3)$, which comes anyhow from a class of the base, \eqref{2ChernclassZ4liscio} is manifestly a linear combination of wedge products of two classes belonging to $H^2(Z_4,\mathbb{Z})=H^{1,1}(Z_4,\mathbb{C}) \cap H^2(Z_4,\mathbb{Z})$, which never have only one leg along the $T^2$ fiber.

To summarize, any Poincar\'e preserving flux $G_4$ in a CY fourfold with SU(4) holonomy must be such that:
\begin{equation}
\int_{Z_4} G_4 \cdot D \cdot D' = 0 \qquad \text{for any two} \quad D, D' \in H^{1,1}(Z_4,\mathbb{C}) \cap H^2(Z_4,\mathbb{Z})\,.
\end{equation}

\emph{Therefore, if $G_4$ is to preserve Poincar\'e invariance in $\mathbb{R}^{1,3}$,  $c_2(Z_4)$ must always be even}, so that $G_4$ can be shifted back to zero.

In order to address this issue, it is convenient to reduce the problem to a more tractable one formulated entirely on the base space of the fibration.

The term $12F^2$ in \eqref{2ChernclassZ4liscio} is \emph{twice} the Poincar\'e dual of a true, perfectly decent 4-cycle, which is a divisor of $B_3$. Indeed, this 4-cycle is described in $Z_4$ by the two equations $X=0$ and $Y=0$, which have class $2F$ and $3F$ respectively. 
Hence, its Poincar\'e dual is an integrally quantized 4-class and thus $12F^2$ is always an even class.

In this way, one is led to analyze a pure class of the base, namely $c_2(B_3)-c_1^2(B_3)\in H^4(B_3,\mathbb{Z})$. In principle, we should integrate this class on all divisors of $B_3$, in order to decide whether or not it is always even. Fortunately, some basic facts from algebraic topology will help us to determine the general answer to this problem. We state the following proposition:
\paragraph{Proposition:}\label{Fact} For any \emph{smooth, complex} manifold $X$ of \emph{complex dimension at most three}, the characteristic class $c_2-c_1^2$ of its tangent bundle is always \emph{even}.
\paragraph{Proof:} It is convenient to reduce the integral 4-class in question to a class modulo 2 and study whether the latter vanishes or not.
Define the \emph{second Stiefel-Whitney class} $w_2$ as \emph{the modulo two reduction of $c_1$}, i.e. $w_2 \equiv c_1\;\textrm{mod}\,2 \in H^2(B_3, \mathbb{Z}_2)$. The modulo 2 reduction of $c_1^2$ is then clearly $w_2^2$ because $c_1\;\textrm{mod}\,2=w_2$, and the quotient homomorphism $q:\mathbb{Z}\rightarrow\mathbb{Z}_2$ induces a homomorphism of cohomology rings under the cup product $\cup$ (which in this case is simply the wedge product). For $c_2$, instead, one notices that, since the manifold is complex, it is equal to the class $\lambda$ apart from the sign, which is killed by the mod 2 reduction. However, as stated in \cite{Witten:1996md}, the mod 2 reduction of $\lambda$, \eqref{lambdadef}, is the so-called \emph{fourth Stiefel-Whitney class} $w_4$. Thus, for complex manifolds\footnote{In general, for a spin$^c$ manifold which is not complex, one has \cite{Witten:1999vg, Sati:2010mj} $w_4=\lambda\;\textrm{mod}\,2=\frac{p_1-\alpha^2}{2}\;\textrm{mod}\,2$, where $\alpha$ is the spin$^c$ class. In particular, if the manifold is complex, then $\alpha=c_1$.}, $c_2\;\textrm{mod}\,2=w_4$. Therefore, one is led to analyze the class:
\be\label{c2menoc1quadro}
c_2-c_1^2\quad\textrm{mod}\;2&=&w_4+w_2^2\quad\in\; H^4(X,\mathbb{Z}_2)\,.
\ee
It is here that certain tools of algebraic topology become useful, namely, the so-called \emph{Steenrod squares} and \emph{Wu classes}. The Steenrod squares are operations in the $\mathbb{Z}_2$-cohomology of a given space $M$ of real dimension $n$, such that $\sq^i:H^k(M,\mathbb{Z}_2)\rightarrow H^{k+i}(M,\mathbb{Z}_2)$. Fortunately, we will not need to define these objects. It turns out that one can extract crucial information out of them just by knowing some of their basic properties and relations. We summarize them below \cite{MS, StongYoshida}:

\begin{enumerate}
\item $\sq^0$ is the identity map.
\item The total Steenrod square respects the cup product, namely:
\[\sq^k(x\cup y)=\sum_{i+j=k}\sq^i(x)\cup\sq^j(y)\,.\]
\item The Wu classes $v_i$ are defined as the unique representatives (by Riesz's theorem) of the functionals (upon integration on $M$) $\sq^i(x)$ for $x\in H^{n-i}(M,\mathbb{Z}_2)$, namely:
\[\sq^i(x)\equiv v_i\cup x\,,\qquad x\in H^{n-i}(M,\mathbb{Z}_2)\,,\qquad0\leq i\leq n\,.\]
\item The total Stiefel-Whitney class equals the total Steenrod square applied to the total Wu class, namely:
\[w_i=\sum_{j=0}^i\sq^{i-j}(v_j)\,.\]
\item They satisfy the so called Wu formula:
\[\sq^i(w_j)=\sum_{t=0}^i\left(\begin{array}{c} j+t-i-1\\ t\end{array}\right)w_{i-t}\,w_{j+t}\,.\]
\item Since $\sq^i(x)=0$ if $x\in H^j(M,\mathbb{Z}_2)$ with $i> j$, then, by definition:
\[v_i=0\qquad\forall\,i>\left[\frac{n}{2}\right]\,.\] 
\end{enumerate}
Using these properties, one can find very easily the expressions of the Wu classes in terms of the Stiefel-Whitney ones. For example, for the first four, one gets: 
\be\label{Wuclasses}
v_1&=&w_1\,,\0\\
v_2&=&w_2+w_1^2\,,\0\\
v_3&=&w_1w_2\,,\0\\
v_4&=&w_4+w_2^2+w_1^4+w_1w_3\,.
\ee
Now, since the manifold at hand, $X$, is complex, it is in particular orientable, so that  $w_1$ of its tangent bundle vanishes. Hence, in the present situation, the fourth Wu class displayed in \eqref{Wuclasses} becomes exactly the reduction modulo 2 of the integral class of \eqref{c2menoc1quadro}:
\be
v_4&=&c_2-c_1^2\quad\textrm{mod}\;2\,.
\ee
However, the last of the properties listed above says that if $\textrm{dim}_{\mathbb{C}}X\leq3$, then $v_i=0$ for all $i\ge4$.\\ This concludes the proof.  \\ $\square$

\paragraph{}Notice that, the hypotheses of this result constitute no restriction on the base $B_3$ of the elliptic fibration, since it is always a K\"ahler manifold, and thus complex. 
In order to prove that this quantity is instead vanishing for $Z_4$, it was crucial for the latter to be elliptically fibered with a globally defined \emph{smooth} Weierstrass representation. 

There are two important conclusions we can draw from this result:

\begin{itemize}
\item Had there been any smooth, elliptically fibered $Z_4$'s with a Weierstrass representation with odd $c_2(Z_4)$, then the flux half-integral quantization would have forced them to break Poincar\'e invariance. Hence, such fourfolds would have been ruled out as F-theory compactifications from their inception.

\item All sources of flux half-integral quantization must be localized around stacks of 7-branes, or other potential sources of singularities of $Z_4$. 
In fact, there is no reason for this result to hold for a generic Calabi-Yau fourfold. 
This is what motivates us to allow for non-Abelian singularities for $Z_4$ in section \ref{SingularCase}. 

\end{itemize}

In the singular case, the computation of the second Chern class by means of the adjunction formula \eqref{ChernclassZ4liscio} will no longer be reliable, thus forcing one to resolve the singularity via a series of blow-up's. Only after the complete resolution can one compute $c_2$ using adjunction. However, the resolved fourfold, although still an elliptically fibered Calabi-Yau (all the resolutions in the series will be crepant), will in general no longer admit a Weierstrass description of the fibration over the whole $B_3$. Hence, it will evade our proposition, and may induce half-integrally quantized fluxes.

As a last comment, it is important to stress that when $\lambda$ of the fourfold is an even class as in this case, the geometric tadpole of F-theory, namely $\chi(Z_4)/24$, is always an integer, as Witten proved using index theorems \cite{Witten:1996md}. Computing the holomorphic Euler characteristic of $Z_4$ by means of the Hirzebruch-Riemann-Roch theorem, one has:
\be
2=\chi_0(Z_4)=\int_{Z_4}\textrm{Td}(Z_4)=\frac{1}{720}\int_{Z_4}3c_2^2(Z_4)-c_4(Z_4)\,,
\ee
where $\textrm{Td}$ is the total Todd-class and, the strict SU(4) holonomy implies that $h^{0,1}=h^{0,2}=h^{0,3}=0$ for $Z_4$. Therefore, for the minimal choice of $G_4$, one gets:
\be
\frac{1}{2}\int_{Z_4}G_4\wedge G_4=\frac{1}{8}\int_{Z_4}c_2^2(Z_4)=\frac{\chi(Z_4)}{24}+60\,,
\ee
which is integral. In this smooth case, the two terms are separately integer; in section \ref{SingularCase} it will be shown that in general, in the singular case, they are separately non-integral (when computed of course on the blown-up Calabi-Yau), but the total tadpole remains the same as here because the index theorem is valid despite the absence of a Weierstrass representation.

\section{The perturbative type IIB perspective}\label{necessaryB}

It has just been found that F-theory on an elliptic Calabi-Yau fourfold with smooth Weierstrass representation always has a quantized G-flux which preserves 4-dimensional Poincar\'e invariance and no other obligatorily present flux that breaks it. 

The aim of this subsection is simply to check these F-theory expectations from the weak coupling limit viewpoint.
We will first review two view-points on the IIB weak coupling limit: Sen's limit, and the tachyon condensation picture of D7-branes as D9/anti-D9 condensates. Subsequently, we will study two potential sources for half-integrally quantized fluxes: O7-planes on non-spin divisors, and O3-planes.
We will find that, in agreement with our results on smooth fourfolds, there are no obligatory half-integrally quantized fluxes.

\subsection{Review of Sen's limit, and tachyon condensation}
One first parameterizes, without loss of generality, the polynomials $f$ and $g$ of eq. \eqref{Kodairaparametr} as follows:
\be\label{Senhetachi}
f&=&-3h^2+\epsilon\eta\,,\0\\
g&=&-2h^3+\epsilon h\eta+\frac{\epsilon^2\chi}{12}\,,
\ee
where $\epsilon$ is a complex constant which drives the weak coupling limit, while $h$, $\eta$ and $\chi$ are sections of $K^{-2}(B_3)$, $K^{-4}(B_3)$ and $K^{-6}(B_3)$ respectively. At leading order in $\epsilon\rightarrow0$ one finds for the discriminant of the elliptic fibration and for the Klein modular invariant the following expressions:
\be\label{discriminanteSen}
\Delta\,\approx\,-9\epsilon^2h^2(\eta^2+h\chi)\qquad\textrm{and}\qquad j(\tau)\,\approx\,\frac{(24)^4}{2}\frac{h^4}{\epsilon^2(\eta^2+h\chi)}\,.
\ee
Thus, in this limit, $g_s=(\IIm\,\tau)^{-1}$ goes to $0$ everywhere on the base except near $h=0$. This locus is just one of the components of the vanishing discriminant locus and it is interpreted as the divisor of $B_3$ wrapped by an O7-plane. For $\epsilon\equiv0$ there is only a recombined, tadpole-canceling 7-brane wrapping this locus. But when $\epsilon$ is not identically zero, a second component of $\Delta=0$ appears, which should be thought of as the divisor wrapped by a D7-brane, compensating the charge of the O7. Thus:
\be\label{WhitneyUmbrella}
\textrm{O7}\,:\,h(\vec{x})=0\qquad\textrm{and}\qquad\textrm{D7}\,:\,\eta^2(\vec{x})+h(\vec{x})\chi(\vec{x})=0\,,
\ee
where $\vec{x}$ are the coordinates of $B_3$. The particular shape of the D7-brane, the so-called \emph{Whitney umbrella} shape, develops a double curve singularity at the intersection with the orientifold and additional pinch point singularities on this curve where also $\chi=0$. Such singularities make many calculations in type IIB, such as the curvature- and gauge-induced D3-brane tadpole, no longer reliable. Addressing this issue is the core of the paper \cite{Collinucci:2008pf}, where two techniques are developed to actually calculate the total D3 tadpole in order to match the curvature-induced contribution with the prediction from the corresponding F-theory fourfolds. In this paper, we will use the method based on Sen's tachyon condensation, in order to discuss the weak coupling limits of F-theory compactifications on both smooth and singular Calabi-Yau fourfolds. 

It is important to stress (since it will be crucial for the practical computations) that it is very natural to think of the Sen limit as a perturbative type IIB string theory on $\mathbb{R}^{1,3}\times X_3$, where $X_3$ is the Calabi-Yau threefold that double covers $B_3$ with branch locus given by the O7 divisor (and the O3-planes). Here, one adds a new coordinate, $\xi$, odd under the orientifold involution, and a new equation
\be\label{doublecoverCY}
\xi^2=h(\vec{x})\,.
\ee
The resulting configuration is a IIB compactification on $X_3$ with an orientifold involution, $\sigma:X_3\rightarrow X_3$, which also takes into account the $\mathbb{Z}_2$-transformation that acts on the fields of the theory. It acts as $\tilde{\sigma}\equiv\sigma\circ(-1)^{F_L}\circ\Omega$, where $(-1)^{F_L}$ changes the sign of the Ramond states of the left-moving sector, while $\Omega$ denotes world-sheet orientation reversal. \\
However, as already stressed, computations on $X_3$ turn out to be problematic due to the singular shape of the recombined D7-brane (Whitney umbrella). In order to avoid this complication, instead of directly working out the resolution of the singular space, a much easier technique will be adopted that is based on the Sen tachyon condensation and used for similar purposes in \cite{Collinucci:2008pf, Collinucci:2008zs, Collinucci:2009uh}.

The strategy is to consider a pair of D9-branes and the pair of corresponding orientifold-image anti-D9-branes with suitable vector bundles on them, which will eventually tachyon-condense and leave the desired configuration of D7. In general, as explained in \cite{Collinucci:2008pf}, it is crucial to have an even number of D9-branes (and consequently of anti-D9's) in order to avoid a discrete anomaly. Indeed, the worldvolume theory of a D3-brane probe placed with its image on the $O7^-$ in the presence of $r$ D9-branes is an $SU(2)$ $\mathcal{N}=1$ gauge theory coupled to $r$ chiral multiplets in the fundamental of $SU(2)$ coming from open strings stretching from the D3 (and its image) to the $r$ D9's (open strings stretching to the image anti-D9's are simply the image strings and thus should not be counted separately as independent). Therefore, if $r$ is odd, one has an odd number of Weyl fermions in the fundamental of $SU(2)$ and this results in a $\mathbb{Z}_2$-anomaly \cite{Witten:1982fp}. In the present situation one has $r=2$. In section \ref{SingularCase}, cases with bigger $r$ will be also discussed, as they will be needed in analyzing the Sen limit of singular F-theory compactifications.

Let $G$ denote the class of the divisor $\xi=0$, where the O7-plane resides.
The D7-brane should wrap a (singular) divisor of class $8G$.\footnote{Throughout this chapter the charge of D-branes will be computed from the point of view of the double cover $X_3$ and therefore it will be twice the physical D-brane charge.} 
Then, we consider the following configuration of $X_3$-filling D9 and anti-D9-branes with gauge bundles:
\begin{table}[ht]\centering
\begin{tabular}{ccccccccc}
$\overline{D9_1}$&&$\overline{D9_2}$&&&&$D9_1$&&$D9_2$\\
$\mathcal{O}(-aG)$&&$\mathcal{O}((a-4)G)$&&&&$\mathcal{O}(aG)$&&$\mathcal{O}((4-a)G)$\;,
\end{tabular}\end{table}\\
where $a$ is some integer. Here, by $\mathcal{O}(n\, G)$ we mean in general the line bundle whose first Chern class is $n\,G$. We will also refer to a polynomial being ``of degree $n\,G$'', whenever it is a section of $\mathcal{O}(n\, G)$. This is compatible with the O7-projection because the anti-D9's are exactly the orientifold images of the D9's. The tachyon $T$ of this configuration is a $2\times2$ matrix-valued section of $E\otimes E$, where $E=\mathcal{O}(aG)\oplus\mathcal{O}((4-a)G)$. The orientifold involution acts on the tachyon as follows:\footnote{The minus sign corresponds to the choice of the $O7^-$; the $O7^+$ involution would have had the plus sign. The transpose, instead, comes from the fact that $\Omega$ exchanges the open string endpoints.}
\be
\tilde{\sigma}^*(T(\vec{x},\xi))&=&-{}^tT(\vec{x},-\xi)\,,
\ee
where $\vec{x}$ are the local coordinates on $B_3$. To survive such an orientifold projection, the tachyon is constrained to take the following form:
\be
T(\vec{x},\xi)&=&\left(\begin{array}{cc}0&\eta(\vec{x})\\-\eta(\vec{x})&0\end{array}\right)\,+\,\xi\left(\begin{array}{cc}\rho(\vec{x})&\psi(\vec{x})\\ \psi(\vec{x})&\tau(\vec{x})\end{array}\right)\,,
\ee
where $\eta$ is the same polynomial of degree $4G$, as in \eqref{WhitneyUmbrella}, while $\rho,\psi,\tau$ are other locally defined polynomials such that $\rho\tau-\psi^2\equiv\chi$ of formula \eqref{WhitneyUmbrella}. Their degrees are $(2a-1,3,7-2a)G$ respectively. In order for these polynomials to be well-defined and not identically zero, $a$ should satisfy the following bounds:
\be\label{semiintegerbound}
\frac{1}{2}<a<\frac{7}{2}\,.
\ee
The D7-brane locus is by definition the locus where the tachyon map fails to be invertible, i.e. where its determinant vanishes:
\be
S_2\,:\;\det(T)\,=\,\eta^2+\xi^2(\rho\tau-\psi^2)\,=\,0\,.
\ee
This reproduces the Whitney umbrella shape defined in \eqref{WhitneyUmbrella}, except for the a priori non-generic form of $\chi=\rho\tau-\psi^2$, which we will discuss next.

\subsection{Non-spin O7-plane}

In the previous section, we saw that the form of $\chi$ is non-generic unless one of the two bounds \eqref{semiintegerbound} for $a$ is saturated (i.e. either $\rho$ or $\tau$ becomes a constant). 

The analysis from \cite{Collinucci:2008zs} reveals that if the first Chern class of the normal (line) bundle of the O7 inside $X_3$ is odd, the bound for $a$ cannot be saturated. Indeed, in order to saturate, say, the lower bound $a=\tfrac{1}{2}$, the class $G$ must itself be even.
Hence, if $G$ is odd, the observed effect would signal the presence of a gauge flux \`a la Freed-Witten which cannot be put to zero: Indeed, this flux would create a superpotential \cite{Martucci:2006ij} that would constrain the transverse moduli of the D7-brane.
Since, by adjunction, $c_1(N_{X_3}O7)=-c_1(O7)=c_1(B_3)|_{O7}$, the oddness of the class of $\xi=0$ just means that the divisor wrapped by the O7 is not spin, which also implies that $B_3$ itself is not spin. 

After reviewing this situation, we will prove that the necessity of such a gauge field can be circumvented by turning on a discrete B-field, consistent with the orientifold projection. This is as expected from F-theory, which, as proven in section \ref{casoliscio}, does not predict the appearance of any 7-brane gauge flux with shifted quantization condition.

This phenomenon can also be deduced from the contribution to the D3-tadpole of this system, whose practical computation will be now sketched for later use. \\
Gauge and gravitational couplings of a Dp-brane induce lower dimensional D-brane charges according to the unique non-anomalous coupling found by Minasian and Moore \cite{Minasian:1997mm}. Here one has D9-branes wrapping the whole Calabi-Yau threefold, which means that their normal bundle is trivial and the $\hat{A}$-genus becomes equal to the Todd class of $X_3$. Thus the contribution to lower dimensional D-brane charge densities due to the $\overline{D9}-D9$ system is:
\be\label{contributoD9}
\Gamma_{D9}=\ch\,([E]-[\bar{E}])\cdot\sqrt{\textrm{Td}(X_3)}=\left(e^{aG}+e^{(4-a)G}-e^{-aG}-e^{(a-4)G}\right)\cdot\left(1+\frac{c_2(X_3)}{24}\right)
\ee
where $\bar{E}$ is the gauge bundle on the anti-D9's and the square brackets denote the K-theory class. Also an Op-plane contributes to lower dimensional D-brane charges via gravitational coupling only (since there is no gauge bundle on it), according to the formula \cite{Morales:1998ux}:
\be\label{WZorientifolds}
S_{WZ}^{Op^\pm}&=&\pm\,2^{p-4}\int_{Op}i^*C\wedge\frac{\sqrt{L\left(\frac{1}{4}TOp\right)}}{\sqrt{L\left(\frac{1}{4}N_{X_3}Op\right)}}\,,
\ee
where $i$ is the embedding of the Op-plane worldvolume in the target space, while $L$ is the Hirzebruch genus. Hence, for the case of the $O7^-$, the first non-trivial induced charge density is the D3-brane one and it is easy to see that it reduces to:
\be\label{contributoO7}
\Gamma_{O7}=\frac{\chi(O7)}{6}\,.
\ee
First of all, the zeroth order term in \eqref{contributoD9} vanishes, indicating that no net charge of D9-brane is left after the tachyon condensation (stated in mathematical terms, this means that on the $\overline{D9}-D9$ system there is a class in $\tilde{K}(X_3)$). Then, the first order term in \eqref{contributoD9} gives the right D7-brane charge, namely $8G$, while the second order term vanishes, compatibly with the fact that D5-branes are projected out by the orientifold involution.\\
The third order term in  \eqref{contributoD9} and the contribution  \eqref{contributoO7} constitute the total induced D3-brane charge, which must be compensated by an equal number of explicitly added D3-sources in order to cancel the D3-brane tadpole. It is possible to single out two different contributions to this charge density:
\be\label{finale}
Q_{D3}&=&Q_{gauge}+Q_{grav}\,.
\ee
The gauge contribution reads:
\be\label{contributogauge}
Q_{gauge}&=&4\left(a-\frac{1}{2}\right)\left(a-\frac{7}{2}\right)G^3\,,
\ee
which manifestly disappears, as it should, when the bound \eqref{semiintegerbound} is artificially saturated (i.e. absence of D7-brane gauge flux). The gravitational contribution, instead, reads:
\be\label{contributograv}
Q_{grav}&=&\frac{29}{2}G^3+\frac{c_2(X_3)}{2}G\,.
\ee
The curvature-induced contribution \eqref{contributograv} should exactly match the geometric tadpole predicted by the F-theory lift of the system (or better twice it, because the image-D3's are counted separately); and indeed it does:
\be
\frac{\chi(Z_4)}{12}=\int_{B_3}30G^3+c_2(B_3)G\,,
\ee
where, analogously to \cite{Sethi:1996es}, the fourth order term of \eqref{ChernclassZ4liscio} has been used together with the fact that the class $F$ integrates to 1 on the generic $T^2$ fiber (i.e. the 0-section is not a multi-section). Using the adjunction formula $c_2(B_3)|_{X_3}=c_2(X_3)-G^2|_{X_3}$, this can be turned into an integral which involves only Calabi-Yau threefold data\footnote{The computation on $X_3$ is actually the only reliable one, since $B_3$ is singular on the O7 locus in the weak coupling limit and sometimes also because of the presence of O3-planes.}:
\be\label{matchingD3tadpole}
\frac{\chi(Z_4)}{12}=\frac{1}{2}\int_{X_3}29G^3+c_2(X_3)G\,,
\ee
where the factor $1/2$ takes into account that $X_3$ is a double cover of $B_3$. This exactly matches the integration on $X_3$ of \eqref{contributograv}. 

Therefore, if things stay like this, it seems impossible to eliminate the gauge contribution \eqref{contributogauge}, with the unexpected consequence of a compulsory presence of D7-brane gauge flux. However, one can turn on a half-integral B-field (with topological type related to $w_2(B_3)$, see below) to cure this problem. 
The details of how such a discrete B-field modifies the interval \eqref{semiintegerbound} making its extrema integral will be now described.

If a B-field is turned on in this type IIB orientifold compactification, this should be done compatibly with the orientifold projection, which acts on it as:
\be
\tilde{\sigma}^*(B(\vec{x},\xi))&=&-\sigma^*B(\vec{x},-\xi)\,,
\ee
$\sigma^*$ being the pull-back map of the target space involution, acting on its cohomology ring.
Hence, in order for the B-field to survive to the projection, one would have to require $B+\sigma^*B=0$. However, since what really matters at the quantum level is the holonomy of the B-field, one is led to impose the weaker condition $B+\sigma^*B\in H^2(X_3,\mathbb{Z})$, namely $\textrm{Hol}(B+\sigma^*B)=1$. One could now wonder which representative in the second de Rham cohomology of $X_3$ should be chosen for the B-field. This, of course, does not matter for closed strings which do not feel large gauge transformations; but for open strings different choices are in general not equivalent, because integral shifts for the $B$ may change their path integral measure. The right trivialization is provided \cite{Bonora:2008hm} by the gauge fields on the D9-branes, in such a way that the combination $B+F$ remains gauge invariant. Actually there is the freedom to choose only one overall trivialization, on one of the four D9's, as the choice for the other gauge fields is constrained by the orientifold involution. Indeed, for example, one can conventionally choose to trivialize the B-field on $D9_1$ by means of the gauge field $F_1$, namely one chooses the couple $(F_1,B)$ on $D9_1$. Then, the involution would imply the couple $(-\sigma^*F_1,-\sigma^*B)$ for the $\overline{D9}_1$. But, since the bulk B-field is the same and these are branes filling the whole target space, the right trivialization on the $\overline{D9}_1$ turns out to be $F_1'=-\sigma^*F_1-B-\sigma^*B$, so that on the $\overline{D9}_1$ one has the couple $(F_1',B)$ (so one has just applied the large gauge transformation $B+\sigma^*B$ to the previous trivialization). Analogously, choose the couple $(F_2',B)$  on the $\overline{D9}_2$, namely the trivialization $F_2'$ on it, which is anyhow fixed by the 7-brane tadpole cancellation $-(F_1'+F_2')=4G$. Then, again applying the large gauge transformation $B+\sigma^*B$, the right trivialization on the $D9_2$ will be $F_2=-\sigma^*F_2'-B-\sigma^*B$, so that on the $D9_2$ one has the couple $(F_2,B)$.\\ 
Now, as opposed to the target space orientifold involution defined in \cite{Collinucci:2009uh}, in the present case the involution just reverses the sign of a coordinate: therefore, its action on cohomology is trivial, namely $\sigma^*$ is the identity map. With the minimal choice $2B=G$, thought of as a class of $X_3$ by pull-back\footnote{On $B_3$, instead, this is exactly what has been said above, namely $2B\;\textrm{mod}\,2=w_2(B_3)$. Hence, $\textrm{Hol}B\in H^2(X_3,\Gamma_2)$, where $\Gamma_2$ represents the subgroup of $S^{1}$ given by the square roots of unity.}, one has the following new configuration:
\begin{table}[ht]\centering
\begin{tabular}{ccccccccc}
$\overline{D9_1}$&&$\overline{D9_2}$&&&&$D9_1$&&$D9_2$\\
$\mathcal{O}(-(a+1)G)$&&$\mathcal{O}((a-4)G)$&&&&$\mathcal{O}(aG)$&&$\mathcal{O}((3-a)G)$\;.
\end{tabular}\end{table}\\
It is important to stress that such gauge freedom was absent in the previous configuration, because 
$\textrm{Hol}B=0$ there, and thus the choice $B=0$ was \emph{canonical}.\\
If one computes the interval for $a$ in this new configuration, one easily finds:
\be\label{newbound}
0\leq a\leq3\,,
\ee
where the extrema are now integral and the bound can be honestly saturated. 

It is useful to verify, as a crosscheck, that the D3 charge computation, based as before on the rational K-theory formula, gives, upon saturation of the bound (i.e. by killing the gauge contribution), precisely the same number as the F-theory geometric tadpole. But now it is crucial to remember that the Chern character of the K-theory class on the $\overline{D9}-D9$ system must be represented by the exponential of the gauge invariant combination $B+F$, that in this particular case turns out to be also quantized (although only in terms of half-integers), because of the constraint of the orientifold projection on the B-field. So in this case $\textrm{exp}(B+F)$ is the only available concept of gauge-induced lower-dimensional D-brane charges and it should enter the Wess-Zumino action of the $\overline{D9}-D9$ system.\\
Therefore, the new contribution due to the D9's and image-D9's will be:
\be\label{contributoD9B}
\Gamma^B_{D9}=\left(e^{(a+\frac{1}{2})G}+e^{(\frac{7}{2}-a)G}-e^{-(a+\frac{1}{2})G}-e^{(a-\frac{7}{2})G}\right)\cdot\left(1+\frac{c_2(X_3)}{24}\right)\,,
\ee
while the contribution of the O7-plane remains obviously the same as in \eqref{contributoO7}. \\
Putting everything together, one finds again vanishing D9's and D5's net charges, the right D7-brane charge ($8G$) and finally $Q_{grav}$ as in \eqref{contributograv}, while for the gauge contribution to the D3-tadpole:
\be\label{contributogaugeB}
Q^B_{gauge}&=&4\,a\left(a-3\right)G^3\,,
\ee
which can manifestly be killed by saturating the new bound \eqref{newbound} and there is no real topological obstruction to choose a vanishing gauge flux on the D7-brane. Notice that in the case the class $G$ is \emph{even} the bound \eqref{semiintegerbound} can be saturated. Consequently, the B-field just introduced becomes integrally quantized and hence one can more easily choose the canonical gauge $B=0$.

Finally note that the M-theory origin of this B-field is \emph{not} $G_4$, because the associated H-field vanishes identically for Freed-Witten anomaly cancellation \cite{Freed:1999vc}. It is rather the holonomy of the $C_3$-field (connection on the membrane 2-gerbe) that generates this $B$ via the F-theory limit.

\subsection{O3-planes}

This paragraph contains some comments about the possible presence of O3-planes in F-theory compactifications, resulting in codimension four singularities of different nature with respect to the non-abelian one discussed in the next section.

First of all, notice that the $\mathbb{Z}_2$-orbifold singularity of $B_3$ represented by the O7-plane is only an artifact of the perturbative limit. Indeed the O7 should really be thought of  as a bound state of two mutually non-local 7-branes placed at a non-perturbatively small distance of the order of $e^{-1/g_s}$ \cite{Denef:2008wq}. Thus the F-theory background, which resolves this distance, does not feel any singularity at all.

A different story, instead, is the one related to other possible fixed loci of the orientifold involution: there can be, in fact, codimension three such loci (points) on the Calabi-Yau threefold, which become  singularities at the quotient. These are interpreted as O3-planes and, since they are singlets of S-duality, they contain no charge for the axio-dilaton and thus the elliptic fiber does not degenerate on them. For this reason, F-theory does not see them and they generate point-like orbifold singularities also on the elliptic Calabi-Yau fourfold. This can be rigorously shown as follows.\\
An O3-plane is present, for instance, whenever the point described\footnote{We remark that here $\xi$ is treated as a \emph{toric} coordinate of $X_3$, and thus not as a polynomial function of the other coordinates.} by $\xi=1$ and all the coordinates of $X_3$  set to zero is fixed under the orientifold involution, due to some projective rescaling\footnote{This is what happens, for example, in the case of the quintic, discussed in \cite{Collinucci:2008zs}.} $\xi\rightarrow\lambda\xi$. Let $h\equiv\xi^2$ the coordinate that substitutes $\xi$ in the quotient $B_3$: due to the fact that $h$ scales with $\lambda^2$, $B_3$ has a $\mathbb{Z}_2$-orbifold singularity in $h=1$. The question is how this singularity is seen by the elliptic fibration via its Weierstrass representation \eqref{Kodairaparametr}. In this situation, \eqref{Kodairaparametr} simplifies to:
\be\label{weierstrassO3}
Y^2=X^3+ah^2XZ^4+bh^3Z^6\,,
\ee 
where the Sen parameterization of $f$ and $g$, \eqref{Senhetachi}, has been used and $a,b$ are constants which take into account the now present dependence on $h$ also of all the other terms in \eqref{Senhetachi}, apart from the first. The two projective rescalings of the four homogeneous coordinates appearing in \eqref{weierstrassO3} are:
\be
(h,X,Y,Z)\sim(\lambda^2h,\lambda^2X,\lambda^3Y,Z)&\Longleftrightarrow&(h,X,Y,Z)\sim(\lambda^2h,X,Y,\lambda^{-1}Z)\,,\label{primoresc}\\
(h,X,Y,Z)&\sim&(h,\mu^2X,\mu^3Y,\mu Z)\,,\label{secondoresc}
\ee
where the first line displays two different and equivalent projective weight assignments for the rescaling of the base coordinates (the second choice is obtained by subtracting the $\mu$-weights from the $\lambda$-weights of the first choice). It is easy to see, then, that as long as $Y\neq0$ and $Z\neq0$, one does not get any orbifold singularity in $h=1$, because any of the two choices of basis for the first rescaling fixes completely the sign freedom of $\lambda$.\\
However, if one among $Y$ and $Z$ vanishes, one immediately gets singularities. Indeed, take for example $Y=0$. Thus, one can fix the $\mu$-rescaling \eqref{secondoresc} by choosing $Z=1$, since, \eqref{weierstrassO3} with $Y=0$ imposes $Z\neq0$; then, by adopting the first choice of basis in \eqref{primoresc}, it is evident that the points $(1,X_i,0,1)$, where $X_i$ ($i=1,2,3$) solve the cubic equation $X^3+aX+b=0$, are fixed under the gauge transformation $\lambda=-1$: Hence these are generically three points with a $\mathbb{Z}_2$-orbifold singularity.\\
Similarly, if $Z=0$, eq. \eqref{weierstrassO3} imposes $X\neq0$ and one can fix it to 1 using part of the $\mu$-gauge \eqref{secondoresc}; then, by choosing  the second basis in  \eqref{primoresc}, it is evident that the points $(1,1,\pm1,0)$ are fixed under the gauge transformation $\lambda=-1$; but this is actually just \emph{one point}, because the residual freedom in the $\mu$-rescaling exchanges $(1,1,1,0)$ with $(1,1,-1,0)$: Hence this is one point with a  $\mathbb{Z}_2$-orbifold singularity.

The presence of O3-planes enters crucially the above discussion as far as the integrality of the gravitational tadpole \eqref{contributograv} is concerned. Indeed, while $c_2(X_3)$ is always even for the proposition proven at the beginning of the present section, the first term is not always integral. However, if O3-planes are absent, $B_3$ is a perfectly regular manifold, and thus one can compute its holomorphic Euler number by means of the Hirzebruch-Riemann-Roch theorem. Assuming proper SU(4) holonomy for $Z_4$, we know that $h^{0,1}(B_3)=h^{0,2}(B_3)=h^{0,3}(B_3)=0$, since, otherwise, holomorphic 1-, 2- or 3-forms would be generated on $Z_4$ via pull-back. Therefore, one has:
\be
1=\chi_0(B_3)=\int_{B_3}\textrm{Td}(B_3)=\frac{1}{24}\int_{B_3}c_1(B_3)c_2(B_3)\,.
\ee 
But since $c_1(B_3)=G$ and, by adjunction, $c_2(X_3)=c_2(B_3)|_{X_3}+G^2|_{X_3}$, one gets:
\be
\frac{1}{2}\int_{X_3}c_2(X_3)G=24+\frac{1}{2}\int_{X_3}G^3\,,
\ee
and thus, using \eqref{contributograv}:
\be
Q_{grav}&=&24\,+15\int_{X_3}G^3\,,
\ee
which is manifestly integral. Note that also its half, namely the physical D3-brane charge, is integral. Indeed, if $B_3$ is smooth, $G^3\,{\rm mod}\,2=w_2^3(B_3)=0$, since the top cohomology group of $B_3$ does not contain torsion. Taking into account formula \eqref{matchingD3tadpole}, this independently confirms that in the smooth case $\chi(Z_4)$ is a multiple of $24$.

This is just a particular manifestation of a more general fact. Indeed, as will be proven below, 
\be\label{fatto2}
\int_{X_3}G^3\;\textrm{mod}\,2&=&\#\,O3\;\textrm{mod}\,2\,.
\ee
Therefore, in the absence of O3-planes one recovers in \eqref{contributograv} the integrality of the tadpole proved before, while if they are present in odd number (like in the quintic example), then the gravitational tadpole is half-integral. However, in the latter case, one should add to it the contribution of the O3's themselves to the D3-brane tadpole, which, as a consequence of eq. \eqref{matchingD3tadpole}, are not seen by the F-theory computation of the geometric tadpole. This is maybe explained by the high codimension (four, as previously seen) of the elliptic fibration singularities generated by the O3-planes. The contribution of $O3^\pm$-planes to the D3-brane charge, as measured from the covering space $X_3$, is (see formula \eqref{WZorientifolds}):
\be\label{contributoO3}
\int_{X_3}Q_{O3}&=&\pm\frac{\#\,O3}{2}\,,
\ee  
which exactly compensates the half-integrality of the first term in \eqref{contributograv}, giving a perfectly integral gravitational tadpole.\\
Eq. \eqref{fatto2} can be proven as follows. Independently of the presence of O3-planes, the Calabi-Yau threefold $X_3$ and the O7-plane are perfectly regular varieties. Thus, applying adjunction, $c_2(O7)=c_2(X_3)|_{O7}+G^2|_{O7}$, and then:
\be\label{euleroO7G3}
\chi(O7)=\int_{O7}c_2(O7)=\int_{X_3}c_2(X_3)G+G^3\,,
\ee
which implies that $\chi(O7)\;\textrm{mod}\,2=\int_{X_3}G^3\;\textrm{mod}\,2$. Calculating the $\mathbb{Z}_2$-equivariant Lefschetz index on $X_3$, we get:
\be\label{equivariantindex}
L_\sigma=\sum_{i=0}^6(-1)^i(b_+^i-b_-^i)=2h^{0,0}+2\,(h_+^{1,1}-h_-^{1,1})+2\,(h_-^{2,1}-h_+^{2,1}+h^{3,0})\,,
\ee
where $b^i$ are the Betti numbers of $X_3$ and the subscript $\pm$ refers to the behavior under $\sigma$ of the corresponding forms. In the second equality of \eqref{equivariantindex} the following information have been used: $h^{1,0}=h^{2,0}=0$, due to the strict SU(3) holonomy of $X_3$; the unique holomorphic three form of $X_3$ is odd under the involution; the volume form of $X_3$ is $\sigma$-invariant, thus the scalar product between $H^{1,1}=H_+^{1,1}\oplus H_-^{1,1}$ and $H^{2,2}=H_+^{2,2}\oplus H_-^{2,2}$ respect the $\pm$ direct decomposition, and hence $h_\pm^{1,1}=h_\pm^{2,2}$; similar situation holds for the group $H^{2,1}$. But, on the other hand $L_\sigma$ is equal to the Euler number of the loci of $X_3$ fixed under $\sigma$:
\be\label{eulerfixed}
L_{\sigma}=\chi(O7)+\#\,O3\,.
\ee 
Therefore, since \eqref{equivariantindex} is manifestly an even number, \eqref{eulerfixed} and \eqref{euleroO7G3} together  imply \eqref{fatto2}.

\paragraph{Remark} To conclude this section, it is important to mention that the integrality of the physical D3-brane charge (i.e. as computed on the base manifold $B_3$) remains an open issue when O3-planes are present. Indeed, it has been proven above that independently of the presence of O3-planes, the D3-brane charge measured from the Calabi-Yau double cover is always integer, but \emph{not} that it is always an even number! It turns out that actually, if the absolute value of the net O3-plane charge, $|n_{O3^+}-n_{O3^-}|$, is odd, then only one of its signs leads to an integral D3-brane charge as measured on $B_3$ (the other choice gives a semi-integer value, which is unacceptable). For example, in the quintic case considered in \cite{Collinucci:2008zs}, one has only one O3-plane, but only if the latter is an $O3^+$ one gets an \emph{integer} number of D3-branes to be added for tadpole cancellation. \\
Therefore the type of the most abundant O3-planes present seems to be a crucial ingredient as far as the integrality of the D3 tadpole is concerned. Along the lines of \cite{Witten:1997bs}, there should be a connection between the appearance of such $O3^+$ and of a B-field in the bulk, which anyway is necessarily present, according to formula \eqref{fatto2}, if the number of O3-planes is odd. Nevertheless, such a conjecture will not be analyzed any further in this paper, and it will be left for a future work.

\section{Singular Calabi-Yau fourfolds}\label{SingularCase}

We will now treat the case of singular elliptic fourfolds, which is obviously more appealing for GUT model building. The results for the smooth case no longer hold, essentially for two reasons: First, by blowing-up the singularity, one looses the Weierstrass representation. Second, the appearance of exceptional divisors alters the elliptic fiber, making the separation between Poincar\'e-preserving and -breaking G-fluxes no longer as clear as in the previous case. Moreover, one should expect any possible odd value of the second Chern class of the blown-up fourfold to be detected on 4-cycles of the exceptional divisors\footnote{In particular on the ones that collapse after blow-down, which are $\PP^1$-fibrations over a divisor of the brane worldvolume $S_2$.}. Indeed, outside of them, the geometry remains the same as in the smooth case; and $c_2$, being quantized, stays constant during the continuous blow-down process. Such odd values can then be related to the possible absence of spin structure on the D7-brane stack. This would correspond to a Freed-Witten-like gauge flux on the D7's with a shifted quantization condition. Such an effect can be verified in the Sen limit, as done before for the smooth case.\\
We will demonstrate all these effects in the next subsection, in a simple toy model in which an $SU(2)$ singularity is forced by hand on a generic (non-singular) divisor of the base $B_3=\PP^3$. We will then conjecture a general pattern based on Fulton's formula of the total Chern class of blown-up manifolds \cite{Fulton,Andreas:1999ng}. \\
In subsection \ref{SpNsing}, we will address the case of general $Sp(N)$ singularities (which of course contains the previous $SU(2)$ case). We will determine the quantization of the gauge flux and give formulae for gauge and gravitational contributions to the D3 tadpole, starting from the $\PP^3$ toy model and then generalizing the results to any $B_3$ and any divisor class of the O7. Moreover, we will point out for which element of the Cartan torus the gauge flux gets a shifted quantization condition. As we will verify in the next sections, the gauge flux corresponding to the affine node of the extended Dynkin diagram of a given singularity always remains integrally quantized. This node is in fact always present (see appendix \ref{toricblowups}), \emph{also} in the smooth situation, and, at least in unitary gauge theories, it represents the U(1)-trace factor whose gauge boson decouples in the non-abelian case.

\subsection{SU(2) singularities}\label{su2singularities}

In this subsection, we will construct an explicit toy model in detail, in which the F-theory elliptic Calabi-Yau fourfold suffers from a non-abelian singularity of the simplest possible Kodaira type, namely I$_2$. For simplicity, we choose the projective threefold $\PP^3$ as a base. Using the Tate classification given in \cite{Bershadsky:1996nh}, we will constraint the complex structure moduli of the fourfold to generate an $SU(2)$ singularity over a non-singular divisor $S_2\subset B_3$. Then, by means of toric methods,  we will blow-up the fourfold $Z_4$, which will suffice to completely resolve the singularity. Finally, we will compute the second Chern class of the resolved fourfold $\tilde{Z}_4$ and compare it to the first Chern class modulo 2 of $S_2$ (i.e. $w_2(S_2)$).

The reader is referred to the appendices for notations. In appendix \ref{toricblowups} we give some mathematical details about toric blow-up's that will be needed in what follows. Moreover, in appendices \ref{KodSU2} and \ref{compresol} several interesting geometrical features of the $SU(2)$ Kodaira singularity and its toric resolution are described in the simple case of a stack of D7's wrapping a toric divisor. The physical picture behind the geometric construction is also provided.

Suppose we place the $SU(2)$ singularity on a generic divisor of $\PP^3$ described by the equation $P_n(x_1,\ldots,x_4)=0$, where $P_n$ is a polynomial of degree $n$. Since such $S_2$ is not in general toric, a trick is necessary to treat it in the same way as in appendix \ref{KodSU2}. It is sufficient to add a new coordinate to the ambient fivefold, $\sigma$, and a new equation, $\sigma=P_n(x_1,\ldots,x_4)$. The ambient sixfold is a toric variety defined by complex coordinates modulo three weighted projective $\mathbb{C}^*$ actions. The projective weight assignments of the blown-up ambient sixfold, and the degrees of the two equations defining the resolved fourfold are written in the following table:
\be\label{pesiZtildesu2}
\begin{array}{ccccccccc|c|c}x_1&x_2&x_3&x_4&\sigma&X&Y&Z&v&\textrm{eq.}\eqref{quadZ}&v\sigma=P_n\\ \hline
1&1&1&1&n&8&12&0&0&24&n\\
0&0&0&0&0&2&3&1&0&6&0\\
0&0&0&0&1&1&1&0&-1&2&0\end{array}\begin{array}{r}\\ \\ \\,\end{array}\\  \0\\
Y^2+a_1XYZ+a_{3,1}\sigma YZ^3=vX^3+a_2X^2Z^2+a_{4,1}\sigma XZ^4+a_{6,2}\sigma^2Z^6\,,\label{quadZ}
\ee
with $a_i$ being polynomial functions of $x_1\ldots x_4$. The ambient toric sixfold is defined in part by these projective rescalings. One must also specify the `excised' points of the space, much like projective space is defined by first excising the origin of a complex space, and then quotienting by the projective rescaling. The following list, referred to as the \emph{Stanley-Reisner ideal}, has as entries, sets of coordinates that cannot vanish simultaneously :
\be\label{SRIdealX6}
\textrm{SR ideal :}\,\left\{x_1x_2x_3x_4\sigma\,;\,x_1x_2x_3x_4v\,;\,XYZ\,;\,\sigma XY\,;\,vZ\right\}\,.
\ee
Notice that on the blown-up Calabi-Yau fourfold $x_1\ldots x_4$ cannot vanish all at the same time because of the first two elements in \eqref{SRIdealX6} and of the equation $v\sigma=P_n$.

The second Chern class of the blown-up elliptic Calabi-Yau fourfold $\tilde{Z}_4$, defined by (\ref{pesiZtildesu2}-\ref{quadZ}), is readily obtained by adjunction: \label{base2classi}
\be\label{c2su2generico}
c_2(\tilde{Z}_4)&=& \underbrace{11F^2+92FH+182H^2}_{c_2(Z_4)}\;+\;\underbrace{(n-28)EH}_{\Delta c_2}\,,
\ee
where $H$ is the hyperplane class of $\PP^3$ (with weights $(1,0,0)$), $E$ is the class of the exceptional divisor $v=0$ (with weights $(0,0,-1)$) and $F$, differently from the notation of section \ref{casoliscio}, represents the class of the section of the fibration $Z=0$ (with weights $(0,1,0)$). Notice that, with this choice of basis, $EF=0$ because $vZ$ is an element of the SR-ideal of the ambient toric sixfold. The first piece in \eqref{c2su2generico}, $c_2(Z_4)$, is exactly the second Chern class of a smooth representative of the pre-blow-up Calabi-Yau fourfold. The second piece, $\Delta c_2$, instead, is the addition due to the blow-up, which of course depends on the exceptional divisor class.
The intersection numbers of $\tilde{Z}_4$ have been calculated with the program SAGE \cite{sage}. 

The geometric tadpole of F-theory turns out to be:
\be\label{geotadpolesu2gen}
\frac{\chi(\tilde{Z}_4)}{24}&=&972-\frac{1}{4}n(n-28)^2\,,
\ee
which is manifestly integral when $n$ is even and a $1/4$-integer if $n$ is odd. In the latter case, therefore, by Witten's argument mentioned in section \ref{casoliscio}, the second Chern class of $\tilde{Z}_4$ must be \emph{odd}. We expect to give rise to a Freed-Witten-like gauge flux on the D7's. When $n$ is odd, the stack of D7-branes wraps a \emph{non-spin} divisor of $\PP^3$ and, consequently, the Freed-Witten anomaly of open strings requires turning on a half-integrally quantized worldvolume flux.

We will show in the next section that the half-integrally quantized flux arising this way is associated to the Cartan generator of $SU(2)$ and not to the $U(1)\subset U(2)$ which eventually gets decoupled (affine node). Indeed, the latter is also present in the non-singular case, for which it was proven in section \ref{casoliscio} that no such half-integral quantization occurs. 

The two pieces of \eqref{c2su2generico} are orthogonal, i.e. $c_2(Z_4)\cdot\Delta c_2=0$ when integrated on $\tilde{Z}_4$. Thus, there is no mixed contribution to the minimal G-flux induced tadpole of F-theory, $-c_2(\tilde{Z}_4)^2/8$. Hence, the non-integrality of the geometric tadpole of $\tilde{Z}_4$ \eqref{geotadpolesu2gen} should be compensated by the compulsory addition of $-(\Delta c_2)^2/8$, which is due to the blow-up. Indeed, we find that
\be\label{TotalTadpole}
\frac{\chi(\tilde{Z}_4)}{24}-\frac{(\Delta c_2)^2}{8}&=&972\;\;=\;\;\frac{\chi(Z_4)}{24}\,.
\ee
This number is \emph{equal} to the geometric tadpole \emph{alone} of the Calabi-Yau fourfold $Z_4$ without singularities! This nice result has the following physical explanation:\\
From the type IIB point of view, the blow-up transition of the fourfold corresponds to a perfectly smooth process. One is not changing the Calabi-Yau threefold, but rather there is a recombination/separation of branes via tachyon-condensation processes (see subsection \ref{boundstates}). These processes are allowed if all the charges are conserved (kinematics) \emph{and} if the D-terms constraints admit the splittings (dynamics). In M-theory, the transition corresponds to moving in the enlarged K\"ahler moduli space of Calabi-Yau fourfolds. However, the duality between F and M-theory (see \cite{Denef:2008wq}) guarantees that the M2 tadpole does not change during the transition. Therefore, finding the same D3-tadpole after the blow-up procedure constitutes a non-trivial consistency check of the whole reasoning\footnote{It is not at all clear a priori why the $c_2$ of the blown-up Calabi-Yau should really contain the physical information about flux quantization, but it is the most natural candidate.}.

Let us end this subsection with an observation\footnote{We thank T. Grimm for pointing out this aspect.}. The 4d vacuum arising after the smooth transition mentioned above cannot be supersymmetric because the minimal $G_4$-flux we chose cannot fulfill the self-duality condition. Indeed, $G_4=\Delta c_2/2$ has a negative square, as can be deduced from eq. \eqref{geotadpolesu2gen}. Therefore in the orbit of vacua spanned by brane recombination/separation processes, which keep the D3-brane charge to the value $972$, the only supersymmetric point is the vacuum corresponding to the smooth F-theory configuration, where we chose a vanishing $G_4$-flux.

\subsection{Sen's limit of the SU(2) configuration}\label{SenSU2config}

To conclude the treatment of the type I$_2$ $SU(2)$ singularity over a generic divisor in $\PP^3$, we will discuss the Sen limit of such configurations, focusing on some interesting features. 

We will work on the Calabi-Yau threefold $X_3= W\PP^3_{1,1,1,1,4}[8]$, given by the octic hypersurface in the shown weighed projective space with homogeneous coordinates $(x_1, \ldots, x_4, \xi)$. This is the double cover of $\mathbb{P}^3$ branched over the zero-locus $h=0$ of a degree $8$ polynomial $h$. The orientifold involution sends $\xi \mapsto -\xi$.

The perturbative\footnote{In appendix \ref{KodSU2}, we explain the difference between the perturbative and non-perturbative realizations of $SU(2)$.} $SU(2)$ corresponds to a D7-brane with coinciding D7-image, which intersects the O7 transversally. The D7-brane and its image (counted separately) carry a total charge of $2nH$, where $n<16$ is the degree of the polynomial defining their worldvolume and $H$ is the hyperplane class of $X_3$. But, since a total D7-brane charge of $32H$ is needed in order to cancel the tadpole (the O7 has charge $4\,H$ in $X_3$), there must be an orientifold invariant Whitney umbrella D7-brane (see \eqref{WhitneyUmbrella}), with the usual $O(1)$ gauge group, carrying charge $(32-2n)H$.
As opposed to the Sen limit of the smooth case, discussed in subsection \ref{necessaryB}, here one needs 4 D9-branes and 4 anti-D9-branes for the K-theory description of the configuration. The tachyon of the system is given by:
\be\label{tachione4x4}
T&=&\left(\begin{array}{cc}\left(\begin{array}{cc}0&\eta\\-\eta&0\end{array}\right)+\xi\left(\begin{array}{cc}\rho&\psi\\ \psi&\tau\end{array}\right)&0\\ 0&\left(\begin{array}{cc}0&P_n\\ -P_n&0\end{array}\right) \end{array}\right)\begin{array}{c}\\,\end{array}
\ee
so that the total tadpole-canceling D7-brane wraps the manifold:
\be
\det\,T=P_n^2(\eta^2+\xi^2(\rho\tau-\psi^2))=0\,.
\ee 
Following the prescription given in \cite{Collinucci:2008pf} to realize the $Sp(1)$ stack of D7-image-D7 on $P_n=0$, with gauge group-breaking flux in the adjoint, the right configuration of D9 and anti-D9-branes is:
\begin{table}[ht]\centering
\begin{tabular}{ccccccc}
$\overline{D9_1}$&&$\overline{D9_2}$&&$\overline{D9_3}$&&$\overline{D9_4}$\\
$\mathcal{O}((n-14)H)$&&$\mathcal{O}(-2H)$&&$\mathcal{O}\left((14-n)H\right)$&&$\mathcal{O}\left(-14H\right)$\\  \\ $D9_1$&&$D9_2$&&$D9_3$&&$D9_4$\\ $\mathcal{O}((14-n)H)$&&$\mathcal{O}(2H)$&&$\mathcal{O}\left((n-14)H\right)$&&$\mathcal{O}\left(14H\right)$\;.
\end{tabular}\end{table}\\
In order to find the above assignments for the Chern classes of the various bundles in the Whitney sum, four independent conditions have been imposed\footnote{The B field here is absent since the degree of the O7 in $X_3$ is even. Therefore the line bundles on the anti-D9's are just the inverses. The general situation is presented in subsection \ref{generalsit}.}:
\begin{enumerate}
\item right total D7-brane charge, i.e. $32H$;
\item degree $n$ for the polynomial $P_n$;
\item D7 of generic Whitney umbrella shape, i.e., for example, $\textrm{deg}\,\tau=0$;
\item gauge flux on the transverse D7 stack as expected, from the F-theory computation, to give rise to the right gauge contribution to the D3 tadpole (see eq \eqref{geotadpolesu2gen}).
\end{enumerate}
Therefore, the first pair of D9's and anti-D9's is responsible for the Whitney umbrella D7-brane, while the second one for the $Sp(1)$ stack. Notice that for consistency one must require $14-n>2$, i.e. $n<12$, otherwise one looses the generic shape of the Whitney umbrella.

From the above D9-$\overline{\rm D9}$ system, one can compute all induced lower D-brane charges.\\
The net D9-brane and D5-brane charges vanish as they should because of the orientifold projection. The D7-brane charge is instead, by construction, the right one to cancel the 7-brane tadpole, namely $32H$.\\
For the D3-brane charge, (see section \ref{necessaryB}), there is the contribution from the D9-$\overline{\rm D9}$ system and one from the O7-plane. Using that the triple intersection number of the hyperplane class $H$ in $W\PP^4_{1,1,1,1,4}[8]$ is equal to $2$, a simple calculation leads to the following expected result:
\be\label{QD3su2gen}
Q_{D3}&=&1944\;\;=\;\;2\times\frac{\chi(Z_4)}{24}\,,
\ee
which coincides with the total tadpole predicted by F-theory. Now, in order to separate in \eqref{QD3su2gen} the gauge contribution from the gravitational one, recall \cite{Denef:2008wq} that the effective D3-brane charge induced by the D7-brane-type fluxes ensuing after the F-theory limit from $G_4$ is:\footnote{The bulk fluxes part is put to zero here because, as said, no shifted quantization rules arise for them.}
\be\label{secondcherncharact}
-\frac{1}{2}\int_{Z_4}G_4\wedge G_4\,=\,\frac{1}{2}\int_{S_2}\textrm{Tr}\,F\wedge F\,=\,\ch_2(F)\,.
\ee
By construction, the gauge flux on the D7 stack is of the form \cite{Collinucci:2008pf}:
\be\label{gaugefluxsu2gen}
F&=&\frac{1}{2}(28-n)\,H\,\left(\begin{array}{cc}1&0\\0&-1\end{array}\right)\,,
\ee
where the two by two matrix is thought of as taking into account the opposite contributions of the D7-brane and of its image. This is a flux in the adjoint of $SU(2)$, aligned with its Cartan generator. The D3-tadpole contribution from the gauge flux in \eqref{gaugefluxsu2gen}, as seen on the double cover threefold, is thus:
\be\label{gaugesmoothD7}
\int_{S_2}\ch_2(F)=\frac{1}{2}\int_{X_3}\frac{1}{4}\,2\,(n-28)^2H^2\cdot nH=\frac{n}{2}\,(n-28)^2\,.
\ee
This corresponds exactly to the amount one expects from the F-theory calculation, namely to $-2\times (\Delta c_2)^2/8$. Consequently, by subtracting it from the total tadpole \eqref{QD3su2gen}, one obtains twice the geometric tadpole \eqref{geotadpolesu2gen}. Notice that in \eqref{gaugesmoothD7}, the integration on $S_2$ is perfectly well-defined, since the worldvolume of the D7-stack is non-singular. Moreover, the flux \eqref{gaugefluxsu2gen} does not induce any D5-brane charge, since it is traceless and hence the first Chern character vanishes. This is again consistent with the orientifold projection\footnote{Strictly speaking, it is $i_\#\ch_1(F)$, where $i$ is the D7 embedding in $B_3$, that is forced to vanish. But in this case, being $F$ restriction of a class of the target, this is equivalent to the requirement $\ch_1(F)=0$.}.\\
It is important to note that, for odd $n$, the half-integrally quantized D7-brane gauge flux \eqref{gaugefluxsu2gen}, (required for Freed-Witten anomaly cancellation of open strings attached to the non-spin D7-stack), cannot be put to zero. Therefore it necessarily breaks $SU(2)$ to its Cartan torus $U(1)$.

\subsection{Bound states via tachyon condensation} \label{boundstates}

As already mentioned in the previous subsection, one should be able to connect the $SU(2)$ configuration just described to one with a single, recombined Whitney umbrella brane carrying all the D7-charge. This is suggested at the kinematic level by the equality of the total D3-brane tadpole \eqref{TotalTadpole}. Indeed, this is possible by applying to the tachyon field \eqref{tachione4x4} a double change of base of the D9's and the anti-D9's. In this subsection, we describe the process of brane recombination first in the unorientifolded case as a warm-up, and then, in the orientifolded case of interest. This discussion is based on \cite{Brunner:2005fv}.

\subsubsection{Unorientifolded case} \label{unorientifoldedbound}

Before directly addressing the case of interest, let us begin by postponing tadpole cancellation and orientifolding. To describe a single D7-brane as a condensation of a D9/anti-D9 brane\footnote{Alternatively, one can think in terms of D4-branes as D6/anti-D6 condensates.} we use an exact sequence of line bundles as follows:
\begin{equation}
\begin{tabular} {c c c c c}
$\cO(-D+F)$ &$\xrightarrow{T}$ &$\cO(F)$& $\xrightarrow{}$& $\cO(F)|_D$\\ 
$\overline{\rm D9}$ & &D9&  & D7 \\
\end{tabular}
\end{equation}
whereby the notation $\cO(\omega_2)$ is the line-bundle of first Chern class $\omega_2$, and $D$ denotes the divisor class on which the D7-brane is defined. This sequence describes a D7-brane on $D$ with a line-bundle $\cL$ with $c_1(\cL) = F-\tfrac{1}{2}\,D$. The third term in the sequence is a skyscraper-like sheaf localized on $D$ (the cokernel of the tachyon map $T$). The tachyon field is then a section of the line bundle:
\begin{equation}
T \in \Gamma \big(\left( \cO(-D+F) \right)^* \otimes \cO(F) \big) = \Gamma \big( \cO(D) \big)\,.
\end{equation}

Suppose now that two D7-branes with divisor and flux data $(D_1, F_1)$ and $(D_2, F_2)$  intersect along a curve. Then, the following brane recombination process can be triggered by a FI term, provided one chooses the right values for the complexified K\"ahler modulus:
\begin{equation}
(D_1, F_1)  + (D_2, F_2) \quad \mapsto \quad (D_1+D_2, \tilde F)\,,
\end{equation}
where the recombined brane resides on a divisor class equivalent to the sum of the two constituent divisor classes. The charge vector (polyform) for a brane $(D, F)$ is in general:
\begin{eqnarray}
\Gamma_1 &=& Q_{D7} + Q_{D5} + Q_{D3}\\
&=& D+\big(F-\tfrac{1}{2}\,D \big)\,D+ \left(\tfrac{1}{24}\,(c_2(X)\,D+4\,D^3)+\tfrac{1}{2}\,F\,D\,(F-D) \right) \nonumber\,,
\end{eqnarray}
whereby the D7-, D5-, and D3-charges are represented by two-, four-, and six-forms, respectively. If we impose conservation of all three types of charges during the recombination process, we arrive at the following unique constraints (modulo redefinitions):
\begin{equation} \label{recombinationconstraints}
F_1-F_2 = D_1\,, \quad \tilde F = F_1,.
\end{equation}

We would like to describe this process in terms of D9/anti-D9 condensation. We start by combining two exact sequences into one as follows:
\begin{equation} \label{tachyonsequence1}
\cO(F_1-D_1) \oplus \cO(F_2-D_2) \xrightarrow{T} \cO(F_1) \oplus \cO(F_2) \xrightarrow{} \cO(F_1)|_{D_1} \oplus \cO(F_2)|_{D_2}\,.
\end{equation}
Then, the entries of $T$ are sections of the following line-bundles:
\begin{equation}
T \in \begin{pmatrix} \cO(D_1) & \cO(D_2+F_1-F_2) \\ \cO(D_1+F_2-F_1) & \cO(D_2)\,. \end{pmatrix} = \begin{pmatrix} \cO(D_1) & \cO(D_2+D_1) \\ \cO & \cO(D_2)\,. \end{pmatrix}
\end{equation}
where in the last equality we used the constraints \ref{recombinationconstraints}. A general Ansatz for the tachyon matrix is then:
\begin{equation} \label{bindingpattern}
T \in \begin{pmatrix} T_1 & \psi_{1,2} \\ \lambda & T_2 \end{pmatrix}\,.
\end{equation}
When the D9's and anti-D9's annihilate, a D7-brane will be left along the locus where $T$ fails to be invertible, i.e. along
\begin{equation}
|T| = T_1\,T_2-\lambda\,\psi_{1,2} = 0\,.
\end{equation}
Here, we see that if $\psi_{1,2}=0$, then the system corresponds to the union of two different branes on $D_1$ and $D_2$, respectively. However, switching on a vev for $\psi_{1,2}$ corresponds to recombining these two intersecting branes into a single, smooth divisor of class $D_1+D_2$.

In order to see this more directly, we will perform basis transformations on the rank two D9-stack, and the rank two anti-D9-stack, respectively:
\begin{align}
R&: \cO(F_1-D_1) \oplus \cO(F_2-D_2) &\rightarrow& \quad \cO(F_1-D_1) \oplus \cO(F_2-D_2)&\,,\\
L&: \cO(F_1) \oplus \cO(F_2) &\leftarrow& \quad \cO(F_1) \oplus \cO(F_2)&
\end{align}
such that the tachyon transforms as
\begin{equation}
T \mapsto T'=L.T.R\,.
\end{equation}
Choosing
\begin{equation} \label{2by2transfo}
 L = \begin{pmatrix} \tfrac{1}{a} & -\frac{T_1}{a\,\lambda}\\  \\ 0 & -\tfrac{1}{b\,\lambda} \end{pmatrix}\,, \quad R = \begin{pmatrix} b& a\,T_2\\  \\ 0 & -\lambda\,a \end{pmatrix}\,, \quad a, b \in \mathbb{C}^*
\end{equation}
we find
\begin{equation}
T' = \begin{pmatrix} 0 && T_1\,T_2-\lambda\,\psi_{1,2} \\ -1 && 0 \end{pmatrix}\,.
\end{equation}
The $(2,1)$ entry is telling us that the first anti-D9 in \ref{tachyonsequence1} with flux $\cO(F_1-D_1)$ is annihilating the second D9 in \ref{tachyonsequence1} with flux $\cO(F_2 = F_1-D_1)$, leaving us with just one non-trivial sequence:
\begin{equation}
\cO(F_1-D_1-D_2) \xrightarrow{T_1\,T_2-\lambda\,\psi_{1,2}} \cO(F_1)\,,
\end{equation}
corresponding to a D7 on $D_1+D_2$ with flux $F_1$.
Note, that we could have predicted the rule \ref{recombinationconstraints} simply by inspecting the off-diagonal elements of $T$, and requiring that at least one of those be a section of the bundle $\cO(D_1+D_2)$.

\subsubsection{Orientifolded case} \label{orientifoldedbound}

We now move on to the orientifolded, D7-tadpole canceling case. For simplicity of notation, let us focus on our working example, the $W\mathbb{P}^4_{1,1,1,1,4}[8]$ geometry. The following $2 \times 2$ tachyon profile
\be \label{generictachyon}
T_2&=&\left(\begin{array}{cc}0&\eta_{16}\\-\eta_{16}&0\end{array}\right)\,+\,\xi\left(\begin{array}{cc}\rho_{24}&\psi_{12}\\ \psi_{12}&\tau_{0}\end{array}\right)\,,
\ee
where the subscripts indicate degrees, corresponds to the Sen limit of a smooth F-theory fourfold. However, we realize that by tuning the entries as follows:
\begin{equation} \label{tunedentries}
\eta_{16} = \tilde \eta_{16-n}\,P_n\,, \quad \rho_{24} = \tilde \rho_{24-2n}\,{P_n}^2\,, \quad \psi_{12} = \tilde \psi_{12-n}\,P_n\,,
\end{equation}
we arrive at a D7-configuration of the following form:
\begin{equation} \label{su2det}
{\rm det} T = {P_n}^2\, \left( \tilde \eta^2 + \xi^2\, (\tilde \rho\,\tau_0 -\tilde \psi^2) \right)\,.
\end{equation}
This corresponds to a configuration with one Whitney umbrella D7-brane of degree $32-2\,n$, and one $SU(2)$-stack on $P_n=0$. On the other hand, from the techniques in \cite{Collinucci:2008pf}, we know that this configuration can be described by a rank four D9/anti-D9 pair, with tachyon of the following form:

\be  \label{su2tachyon}
T_4&=&\left(\begin{array}{cccc} \xi\,\tilde \rho_{24-2\,n}&\tilde \eta_{16-n}+\xi\,\tilde \psi_{12-n} &  \frac{\xi\,P_n\, \tilde\rho_{24-2\,n}}{\lambda_0}& \lambda_0 \\ 
-\tilde \eta_{16-n}\,+\xi\,\tilde \psi_{12-n}&-\xi\,\tau_0 & 0&0 \\ \\
\frac{\xi\,P_n\, \tilde\rho_{24-2\,n}}{\lambda_0} & 0 &0 &P_n \\ 
-\lambda_0 & 0& -P_n&0\end{array} \right)\,.
\ee
This describes a Whitney umbrella brane on the upper-left $2 \times 2$ block, an $SU(2)$ stack on the lower-right block, and two off-diagonal `binding blocks' whose degrees will be justified shortly. $T_4$ has the same determinant \eqref{su2det}. Note, that this matrix is essentially the same as \eqref{tachione4x4}, except that it also has off-diagonal entries. These entries correspond to bifundamental fields which, when they acquire non-trivial vev's, will bind the Whitney brane to the $SU(2)$ stack. 

We would like to use the technology described in \ref{unorientifoldedbound} to connect $T_4$ back to $T_2$, in order to prove that they are physically equivalent. 

To keep the total D7-charge equal to $32\,H$, we impose that $T_4$ be a map $T_4: \bar E \mapsto E$, where
\begin{equation}
E = \cO(14-n) \oplus \cO(2) \oplus \cO(a) \oplus \cO(-a+n) \,, \quad a \in \mathbb{Z}\,.
\end{equation}
In order for this system to connect to the original tachyon
\begin{equation}
T_2: \bar F \mapsto F = \cO(14) \oplus \cO(2)\,,
\end{equation}
we impose $a = 14$. By reshuffling the line-bundles, we can view the rank four bundle $E$ as 
\be E  = F \oplus \cO(14-n) \oplus \cO(n-14)\,.
\ee
If we rewrite $T_4$ in the basis:
\begin{eqnarray} 
T_4: \cO(-2) \oplus \cO(14-n) \oplus  \cO(n-14) \oplus \cO(-14)  \nonumber \\ \qquad \mapsto 
\cO(14-n) \oplus \cO(14) \oplus \cO(2)  \oplus \cO(n-14)
\end{eqnarray}
then it takes the following form:
\be  
T_4&=&\left(\begin{array}{cccc} \tilde \eta+\xi\,\tilde \psi &\lambda & \xi\,\tilde\rho & \frac{\xi\,P_n\,\tilde\rho}{\lambda} \\
0&P_n  &\frac{\xi\,P_n\,\tilde\rho}{\lambda}  & 0 \\
 - \xi\,\tau & 0 & -\tilde \eta+\xi\,\tilde \psi & 0 \\
   0 & 0 &  -\lambda & -P_n \end{array} \right)\,,
\ee
where we drop the degrees for compactness. If we inspect the $2 \times 2$ in the upper-left and lower-right corners, we recognize the pattern of \eqref{bindingpattern}. We can now apply transformations of the form \eqref{2by2transfo} to act on the lower and upper block separately:
\begin{align}
 L_l &= \begin{pmatrix}1 & 0 & 0 & 0\\ 
 0&1&0&0\\
 0&0& 1 & \frac{-\tilde \eta+\xi\,\tilde \psi}{\lambda}  \\ \\
 0&0& 0 & -\tfrac{1}{\lambda} \end{pmatrix}\,, \quad 
 &R _l = \begin{pmatrix}1 & 0 & 0 & 0\\ 
 0&1&0&0\\
 0&0& 1 & P_n 
 \\0&0& 0 & -\lambda \end{pmatrix}\,, \\
 L_u &= \begin{pmatrix} -1 & 0 & 0 & 0\\ 
 P_n &-\lambda &0&0\\
 0&0& 1 & 0 \\  
 0&0& 0 & 1 \end{pmatrix}\,, \quad 
& R _u = \begin{pmatrix}1 & 0 & 0 & 0\\
 -\frac{\tilde \eta+\xi\,\tilde \psi}{\lambda}&\frac{1}{\lambda}&0&0\\
 0&0& 1 & 0 \\ 
 0&0& 0 & 1 \end{pmatrix}\,, 
 \end{align}
After performing these transformations and bringing everything back to the original basis:
\begin{equation}
T_4 : \cO(n-14) \oplus \cO(-2) \oplus \cO(-14) \oplus \cO(14-n) \mapsto \cO(14-n) \oplus \cO(2) \oplus \cO(14) \oplus \cO(n-14)\,,
\end{equation}
We have

\begin{equation}
T_4 = \left(\begin{array}{cccc} -\xi\,\tilde\rho & 0 & 0 & -1 \\
0& -\xi\,\tau  & P_n\,(-\tilde \eta+\xi\,\tilde\psi) & 0 \\
  0 & P_n\,(\tilde \eta+\xi\,\tilde\psi) & -\xi\, {P_n}^2\,\tilde\rho & 0 \\
   1 & 0 &  0 & 0 \end{array} \right)\,.
\end{equation}
To make this result more appealing, we perform one more worthwile basis transformation such that 
\begin{equation}
T_4 : \cO(-14) \oplus \cO(-2) \oplus \cO(n-14) \oplus \cO(14-n) \mapsto \cO(14) \oplus \cO(2) \oplus \cO(14-n) \oplus \cO(n-14)\,,
\end{equation}
Now the tachyon profile takes the following form:
\begin{equation}
T_4 = \left(\begin{array}{cccc} -\xi\,\tilde\rho\,{P_n}^2 & P_n\,(\tilde \eta+\xi\,\tilde\psi) & 0 & 0 \\
P_n\,(-\tilde \eta+\xi\,\tilde\psi) & -\xi\,\tau  & 0 & 0 \\
  0 & 0 & -\xi\,\tilde\rho & -1 \\
   0 & 0 &  1 & 0 \end{array} \right)\,.
\end{equation}

We can immediately identify the upper-left block as a Whitney brane of degree $32$, with its entries tuned as stated in \eqref{tunedentries}. The lower-right block has constant determinant one, which means that it is an isomorphism between the D9-stack with $\cO(14-n) \oplus \cO(n-14)$ and its image anti-D9-stack. Hence, the lower-right block physically annihilates, leaving us with one recombined Whitney umbrella. This transformation goes one step further in showing that $T_2$ and $T_4$ are physically connected than just checking charge conservation.  
\vskip 2mm
This can be generalized to construct a rank $N$ D7-stack on $P_n=0$ with a Whitney brane of degree $32-2\,N$. Starting with the Whitney brane of degree $32$ made of a D9-stack with $\cO(2) \oplus \cO(14)$, and its image anti-D9-stack, we simply tag along pairs as follows:
\begin{eqnarray} \label{pretransforpairing}
\underset{\rm Whitney \ of \ deg \ 32}{\underbrace{\cO(2) \oplus \cO(14)}} &&\oplus \underset{\rm trivial}{\underbrace{\cO(14-n) \oplus \cO(n-14) }} \\ &&\oplus \underset{\rm trivial}{\underbrace{\cO(14-2\,n) \oplus \cO(2\,n-14) }} \oplus \ldots \oplus \underset{\rm trivial}{\underbrace{ \cO(14-N\,n) \oplus \cO(N\,n-14) }} \nonumber
\end{eqnarray}
and do the same for the anti-D9's.
Adding these pairs is a trivial operation since their respective tachyon block will be of the form 
\begin{equation}
\left(\begin{array}{cc}  \cdot & -1 \\
  1 & 0 \end{array} \right)\,,
\end{equation}
which is an isomorphism. Then, the next step is to perform a basis transformation analogous to the one performed for the $Sp(1)$ case. Instead of doing this explicitly, we can understand qualitatively what such transformations do. Before the transformation, we view the branes as coming in pairs as indicated in \eqref{pretransforpairing}. After the transformation, the branes will be effectively paired as follows:
\begin{eqnarray} \label{posttransforpairing}
\underset{\rm Whitney \ of \ deg \ 32-2n}{\underbrace{\cO(2) \oplus \cO(14-N\,n)}} &&\oplus \underset{\rm deg \ n \ brane }{\underbrace{\cO(14) \oplus \cO(n-14) }} \\ &&\oplus \underset{\rm deg \ n \ brane }{\underbrace{\cO(14-n) \oplus \cO(2\,n-14) }} \oplus \ldots \oplus \underset{\rm deg \ n \ brane }{\underbrace{ \cO(14-(N-1)\,n) \oplus \cO(N\,n-14) }} \nonumber\,.
\end{eqnarray}
The very first pair (with its image anti-D9 pair) corresponds to the remaining Whitney brane of degree $32-n$, and each of the following $N$ pairs (with their respective anti-D9 pairs) will give rise to a $2 \times 2$ block in the tachyon profile, corresponding to $N$ branes on $P_n=0$.

We will use this pattern of line bundles when we deal with the general structure of Sp$(N)$ stacks later on in \ref{whitneysum} and \ref{generalwhitneysum}.

\subsection{General pattern for the SU(2) configuration}

This section shows how gauge flux quantization in $SU(2)$ singular F-theory compactifications can be described for general base $B_3$. The argument is essentially based on Fulton's formula for the total Chern class of a blown-up manifold \cite{Fulton,Andreas:1999ng}.

Suppose one characterizes the original Calabi-Yau fourfold $Z_4$ as a (singular) hypersurface inside a five-dimensional ambient space $M_5$. In any case, the singularity is placed on a codimension three locus, $S_2\in M_5$, which is itself a perfectly regular surface. Now, one can perform the blow-up of $S_2$ as submanifold of $M_5$, and one has the following commutative diagram:
\be\label{commutativo}
\begin{CD}
E@>j>>\tilde{M}_5\\
@VV{g}V  @VV{f}V\\
S_2@>i>>M_5
\end{CD}
\ee
where $E$ is the exceptional divisor, fibered over $S_2$, and $\tilde{M}_5$ is the blown-up ambient fivefold. The singularity is of multiplicity $2$ in $Z_4$. Therefore:
\be
c_1(N_{\tilde{M}_5}\tilde{Z}_4)=f^*c_1(N_{M_5}Z_4)-\textrm{mult}_{Z_4}(S_2)\cdot E|_{\tilde{Z}_4}=f^*c_1(N_{M_5}Z_4)-2\,E|_{\tilde{Z}_4}\,,
\ee
where the blow-down map restricted to $\tilde{Z}_4$, i.e. $f|_{\tilde{Z}_4}:\tilde{Z}_4\rightarrow Z_4$, has been implicitly used. In this case, Fulton's formula reads \cite{Andreas:1999ng}:
\[
c_2(\tilde{M}_5)=f^*c_2(M)-j_\#g^*(2c_1(S_2)+c_1(N_{M_5}S_2))\,,
\]
where $j_\#$ is the Gysin map in cohomology induced by the embedding $j$. By adjunction and commutativity of diagram \eqref{commutativo} one gets:
\be\label{Curioformula}
c_2(\tilde{M}_5)=f^*c_2(M)-E\,f^*c_1(M_5)-j_\#g^*c_1(S_2)\,.
\ee
The second Chern class of the blown-up Calabi-Yau fourfold $\tilde{Z}_4$ is easily recovered, again by adjunction, as the restriction of \eqref{Curioformula}.\\
The first term in the right hand side of \eqref{Curioformula} represents the second Chern class of the pre-blow-up ambient fivefold and gives rise to the second Chern class of the originally smooth Calabi-Yau fourfold. All the rest, instead, give rise to what has been called $\Delta c_2$ in subsection \ref{su2singularities}, i.e. the crucial additional term due to the blow-up process. Na\"ively using adjunction for $S_2\subset Z_4\subset M_5$, one has, in the basis of section \ref{SingularCase}, $c_1(M_5)=6(c_1(B_3)+F)$ and $c_1(S_2)=F+c_1(B_3)-\PD_{B_3}(S_2)$. Therefore, formula \eqref{Curioformula}, after restriction to $\tilde{Z}_4$,  can be rewritten as follows:
\be\label{formulagenerale}
c_2(\tilde{Z}_4)&=&\underbrace{11F^2+23c_1(B_3)F+11c_1^2(B_3)+c_2(B_3)}_{c_2(Z_4)}+\underbrace{E\,\PD_{B_3}(S_2)-7Ec_1(B_3)}_{\Delta c_2}\,,
\ee
where the general fact $EF=0$ has been used. $c_1(B_3)$ must be understood as $f^*\pi^*c_1(B_3)$ with $\pi:Z_4\rightarrow B_3$ and $E\,\PD_{B_3}(S_2)$ as $j_\#g^*\PD_{B_3}(S_2)$, while $F$ is meant as pulled back by $f$ to $\tilde{Z}_4$.

Formula \eqref{formulagenerale} gives the right result \eqref{c2su2generico} in the case $B_3=\PP^3$, discussed in subsection \ref{su2singularities}. It also gives the general structure  of the second Chern class of the blown-up fourfold for any $B_3$ in terms only of the Chern classes of the base and of the class of the brane worldvolume.

To extend this formula to worse singularities, one should in principle iterate the toric blow-up procedure introducing as many new lattice vectors as required for the complete resolution of the given singularity (see \cite{Bershadsky:1996nh} for a list of such vectors). 

Rather than attempting to perform such iteration, in the next section, we will provide the general expression for the gauge field and the tadpole contributions purely in the type IIB context for the entire series of ${\bf C_N}$ gauge algebras (i.e. symplectic groups $Sp(N)$). We will obtain our results by analyzing the Sen weak coupling limit of such F-theory configurations. The $SU(2)$ results of this subsection will be recognizable by simply putting $N=1$ ($Sp(1)=SU(2)$).

\subsection{Sp(N) singularities}\label{SpNsing}

Before embarking in the full generalization of the D7-brane gauge flux quantization rules for the $Sp(N)$ family of F-theory configurations, it is instructive to first present in detail the case of a $Sp(2)$ singularity on a toric degree one divisor of $\PP^3$ (for example, $x_1=0$).

Resolving the $Sp(2)$ singularity (Kodaira type I$^{ns}_4$) requires adding to the fan of the ambient fivefold $M_5$ the \emph{two} additional vectors \cite{Bershadsky:1996nh} $v_1=(1,1,1,0,0)$ and $v_3=(1,2,2,0,0)$, where the first one is the only one needed for Sp$(1)$.
It is easy to deduce the following projective weights for the various homogeneous coordinates in the game:
\be\label{pesiSp2}
\begin{array}{cccccccrr|ccc}x_1&x_2&x_3&x_4&X&Y&Z&v_1&v_3&&\textrm{proper transform}&{}\\ \hline
1&1&1&1&8&12&0&0&0&&24\\
0&0&0&0&2&3&1&0&0&&6\\
1&0&0&0&1&1&0&-1&0&&2\\
0&0&0&0&1&1&0&1&-1&&2\end{array}\begin{array}{r}\\ \\ \end{array}
\ee 
Therefore, the blow-up generated by $v_3$ is along the codimension three locus in $\tilde{M}_5$ given by $X=Y=v_1=0$, which has multiplicity 2 in the proper transform after the first blow-up. The final proper transform, whose weight assignments are displayed in \eqref{pesiSp2}, has the following form:
\be\label{proptranssp2}
&Y^2+a_1(x_1v_1v_3,x_i)XYZ+a_{3,2}(x_1v_1v_3,x_i)x_1^2v_1YZ^3&=v_1v_3^2X^3+a_2(x_1v_1v_3,x_i)X^2Z^2+\0\\&+a_{4,2}(x_1v_1v_3,x_i)x_1^2v_1XZ^4+a_{6,4}(x_1v_1v_3,x_i)x_1^4v_1^2Z^6&\qquad\qquad\qquad i=2,3,4\,,
\ee
The Stanley-Reisner ideal of the blown-up ambient toric variety is:
\be
\textrm{SR ideal :}\,\left\{x_1x_2x_3x_4\,;\,XYZ\,;\,x_1XY\,;\,x_2x_3x_4v_1\,;\,v_1Z\,;\,v_1XY\,;\,x_2x_3x_4v_3\,;\,v_3Z\,;\,x_1v_3\right\}.
\ee 
From the last element one readily deduces that the exceptional divisor ensuing from the second blow-up, $E_3$, does not intersect the affine node $x_1=0$ of the resolved fiber. This means that its corresponding Cartan node must be the one of the Dynkin diagram of $Sp(2)$ which is invariant under the non-split monodromy. In other words it is the middle node of ${\bf A_3}$ invariant under the symmetry whose folding gives rise to the $Sp(2)$.

After having gauge fixed all the non-zero coordinates, the exceptional divisor $E_1$ is given by the following genus zero \emph{quadratic} curve fibered over $S_2\simeq\PP^2_{x_i}$:
\be
Y^2+a_1(x_i)Y-a_2=0\,.
\ee
Analogously, the exceptional divisor $E_3$ is given by a fibration of the irreducible \emph{quadratic} $\PP^1$:
\be
Y^2+a_1(x_i)XY+a_{3,2}(x_i)v_1Y=a_2(x_i)X^2+a_{4,2}(x_i)v_1X+a_{6,4}(x_i)v_1^2\,.
\ee
Finally, the affine curve is linear, as expected (see appendix \ref{KodSU2}), and it is given again by formula \eqref{affinenode} with $v=v_1$.
By intersecting $E_1$ with $E_3$ one obtains the usual $\mathbb{Z}_2$-fibration over $S_2$, with fiber given by the two points:
\be
Y_{\pm}=\frac{-a_1\pm\sqrt{a_1^2+4a_2}}{2}\,.
\ee
The $\mathbb{Z}_2$-fiber degenerates over the O7, $h\equiv a_1^2+4a_2=0$. Those two points can swap over along closed paths on the base. The same happens to the intersection between $E_1$ and the affine node. All this corresponds to the exchange of the two external nodes of ${\bf A_3}$, or, put differently, to a pairwise exchange of the four D7's realizing the $SU(4)$ gauge group.

The second Chern class of the blown-up fourfold $\tilde{Z}_4$ is:
\be
c_2(\tilde{Z}_4)&=&c_2(Z_4)+\Delta c_2\,,
\ee
where the first term is the usual one of the smooth case (first piece in \eqref{c2su2generico}), while  the additional term reads:
\be
\Delta c_2&=&-H(27E_1+52E_3)\,.
\ee
The intersection numbers have been computed by means of SAGE \cite{sage} and the results are the following:
\be\label{risultatisp2}
\frac{\chi(\tilde{Z}_4)}{24}&=&\frac{1267}{2}\0\\
\frac{(\Delta c_2)^2}{8}&=&-\frac{677}{2}\,.
\ee
By subtracting these two contributions, one obtains the usual D3-brane total tadpole $n_{D3}=972$: This constitutes a non-trivial check of the validity of the results.\\

It is now time to interpret these quantities from the weak coupling type IIB perspective, via Sen's limit procedure. This will lead us to the correct expression of the Freed-Witten gauge flux on the D7 stack. From \eqref{risultatisp2}, one deduces that $c_2(\tilde{Z}_4)$ is odd. Therefore, a gauge flux on the stack of four D7-branes is expected, which is quantized in terms of half-integers.

In order to realize a supersymmetric configuration of two D7-branes and two image-D7-branes on top of each other, wrapping the toric divisor $x_1=0\,\subset\,\PP^3$, transverse to the O7-plane, one needs six D9 branes and six anti-D9-branes. By requiring conditions analogous to the ones of the $SU(2)$ case, and following the procedure explained in \ref{orientifoldedbound}, one ends up with the following Whitney sum of line bundles on the D9-branes (with the conjugate bundles on the anti-D9's):
\begin{table}[ht]\centering
\begin{tabular}{ccccccccccc}
$D9_1$&&$D9_2$&&$D9_3$&&$D9_4$&&$D9_5$&&$D9_6$\\ $\mathcal{O}(12H)$&&$\mathcal{O}(2H)$&&$\mathcal{O}\left(-13H\right)$&&$\mathcal{O}\left(14H\right)$&&$\mathcal{O}\left(-12H\right)$&&$\mathcal{O}\left(13H\right)$\;.
\end{tabular}\end{table}\\
The tachyon field will be a $6\times6$ matrix with determinant:
\be
\textrm{det}\,T&=&x_1^4(\eta^2+\xi^2(\rho-\psi^2))\,,
\ee
where the separation of the two D7-brane components is evident, namely, the $Sp(2)$-stack and the singular Whitney umbrella D7-brane.\\ 
From the above D9-$\overline{\rm D9}$ system one readily computes the various lower dimensional D-brane charges: The D9 and D5-brane charges vanish as they should; the D7 charge is the right one for the tadpole, namely $32H$. Finally, the total D3-brane charge is again equal to $1944$, which confirms that this configuration is connected to the smooth elliptic fourfold on $\PP^3$ by brane recombination/separation processes. \\
The gauge flux induced by tachyon condensation on the D7 stack turns out to be of the following form:
\be\label{CartanDirectionFlux}
F&=&\frac{1}{2}H\left(\begin{array}{cccc}27&0&0&0\\0&25&0&0\\0&0&-27&0\\0&0&0&-25\end{array}\right)\,=\,\frac{1}{2}H(27C_1+25C_3)\,,
\ee
where $C_1=\textrm{diag}(1,0,-1,0)$ and $C_3=\textrm{diag}(0,1,0,-1)$ are the two Cartan matrices of $Sp(2)$. In particular, $C_1$ corresponds to the Cartan node already present in the $SU(2)$ case, as it has the same coefficient in front (see \eqref{gaugefluxsu2gen}). Moreover, we see that such a gauge flux in the adjoint of $Sp(2)$ is quantized in terms of half integers along both the chosen Cartan directions, and it necessarily breaks $Sp(2)$ to its Cartan torus\footnote{More precisely, this is the Cartan torus generated by $C_1,C_3$. Strictly speaking, the gauge flux  \eqref{CartanDirectionFlux}\label{GaugeBreaking} still preserves the $Sp(1)$ factor with orthogonal Cartan generator in the Killing metric.} $U(1)^2$. Finally this gauge field generates exactly the gauge contribution to the D3-brane tadpole expected by the F-theory computation:
\be
\int_{S_2}\ch_2(F)&=&677\,,
\ee
as one can see from formula \eqref{risultatisp2}.

It is now very easy to guess the behavior of higher rank symplectic singularities. The general formulae for any $Sp(N)$ will be given in the following directly for a D7 stack wrapping a divisor of $\PP^3$ of any degree $n$. Notice, however, that $N$ cannot be arbitrarily large: Indeed, for $N\ge6$, the order of zero of the discriminant at that singularity is greater than 10, resulting in the loss of triviality of the canonical bundle of the fourfold.

Let $Sp(0)=1$ by convention, which corresponds to the smooth case of section \ref{casoliscio}. Sen's limit for the F-theory $Sp(N)$ configuration on $P_n=0\subset\PP^3$ can be built from $2N+2$ D9-branes and $2N+2$ anti-D9-branes. Following the procedure in \ref{orientifoldedbound}, we deduce that the gauge bundle on the D9's consists of the following Whitney sum of line bundles (with the conjugate bundles on the anti-D9's):
\be\label{whitneysum}
\mathcal{O}((14-nN)H)\oplus\mathcal{O}(2H)\bigoplus_{i=1}^N\left[\mathcal{O}((in-14)H)\oplus\mathcal{O}((14-(i-1)n)H)\right]\,.
\ee 
The tachyon field is a $(2N+2)\times(2N+2)$ matrix with determinant:
\be
\textrm{det}\,T&=&P_n^{2N}(\eta^2+\xi^2(\rho-\psi^2))\,.
\ee
Consistency requires $nN<12$. One can verify that the Ansatz \eqref{whitneysum} fulfills, upon tachyon condensation, the requirements of vanishing of D9 and D5-brane charge and of right tadpole-canceling D7-brane charge. Moreover, a straightforward computation shows that it also predicts the right total D3-brane tadpole, namely $1944$ in the double cover, as found from the F-theory perspective. This family of configurations is therefore entirely connected to the smooth case by means of a change of basis performed on the tachyon, as expected from \ref{orientifoldedbound}.

The gauge flux induced on the stack of $N$ D7's plus $N$ anti-D7's turns out to be of the following form:
\be\label{flussogauge}
F=\frac{1}{2}H\,\sum_{i=1}^N\,(28-n(2i-1))\,C_{2i-1}\qquad\textrm{with}\qquad (C_{2i-1})_{jk}=\delta_{ij}\,\delta_{ik}-\delta_{i+N,j}\,\delta_{i+N,k}\;,
\ee
where $C_{2i-1}$ are the $N$, $2N\times 2N$ Cartan matrices of $Sp(N)$. The contribution of such a gauge field to the D3-brane tadpole then reads:
\be\label{tadpgaugegen}
\textrm{Tadpole}|_{\textrm{gauge}}&=&\frac{nN}{2}\left[(28-nN)^2+n^2\,\frac{N^2-1}{3}\right]\,,
\ee
while the contribution due to gravitational interactions obviously reads:
\be\label{tadpgravgen}
\textrm{Tadpole}|_{\textrm{grav}}&=&1944-\frac{nN}{2}\left[(28-nN)^2+n^2\,\frac{N^2-1}{3}\right]\,.
\ee
We have checked this general formula numerically also for the cases with $Sp(3)$ and $Sp(4)$ on SAGE, finding complete agreement.

It is clear that, when $n$ is even, namely the D7 stack is wrapping a \emph{spin} manifold inside $\PP^3$, regardless of the value of $N$, \eqref{flussogauge}, \eqref{tadpgaugegen} and \eqref{tadpgravgen} are all integral quantities. In particular, the gauge flux on the D7-branes is integrally quantized and there is no topological obstruction in putting it to zero.\footnote{Notice however that by doing so one unavoidably changes the total induced D3-brane charge. This violates the kinematical constraint necessary to connect by physical processes this family of singular configurations to the smooth one.} On the other hand, if the stack is not spin (if $n$ is odd) then \eqref{flussogauge} implies that a gauge flux on it is generated to compensate for its Freed-Witten anomaly. The quantization of this flux gets shifted by \emph{half}-integers along all the chosen Cartan directions of the gauge group $Sp(N)$. This unavoidable flux breaks $Sp(N)$ to the Cartan torus $U(1)^N$ parametrized by $C_{2i-1}$. However, as already stressed for the $Sp(2)$ case (see footnote \ref{GaugeBreaking}), the $SU(N)$ gauge group orthogonal to the Cartan direction \eqref{flussogauge} survives, so that the group $SU(N)\times U(1)$ is left unbroken by the half-quantized flux. This matches with the expectations from many models available in the literature (see for example \cite{Blumenhagen:2008at}).

To summarize, in the physical transition from a single D7-brane to an $Sp(N)$-stack plus a `remainder' brane, a flux is created on the former stack.  If the D7 stack wraps a non-spin manifold, this gauge field is already sufficient to make the path integral measure of open strings attached to the D7 stack well-defined \cite{Freed:1999vc}. Hence, no further topological obstruction is allowed. This argument rules out the possibility of having  gauge bundles without vector structure, which would imply \cite{Kapustin:1999di} the presence of a non-trivial 't Hooft magnetic flux and, consequently, a new ambiguity of the measure.
Finally, the gauge contribution to the D3 tadpole \eqref{tadpgaugegen} (and clearly also the gravity one, \eqref{tadpgravgen}) is an integer even in the non-spin case, as long as the rank of the original gauge group, $N$, is even, in agreement with the results found from the F-theory perspective.\footnote{Recall that here one is counting the D3-brane charge contributions from the point of view of the Calabi-Yau threefold double cover of $\PP^3$, so its value is \emph{twice} the one found in F-theory.}

\subsection{General pattern for Sp(N) configurations}\label{generalsit}

The only thing left to discuss as far as $Sp(N)$ singularities are concerned, is the generalization of the above formulae to any base $B_3$ of the elliptic fibration and to any degree of the O7-plane as a divisor of the base (in the toy model example it was of degree 4). An additional complication arises here because, as discussed in subsection \ref{necessaryB}, when the O7-plane has an odd degree, a non-trivial B-field is required.
Recalling the construction of the double cover Calabi-Yau threefold over the base $B_3$, the Chern class of the normal bundle of the O7-plane in $B_3$ is equal to the one of the anti-canonical bundle of $B_3$ itself. Therefore, it is useful to define:
\begin{align}
G&\equiv c_1(B_3) = \tfrac{1}{2}\,\PD_{B_3}(\{h=0\})\,\in\,H^2(B_3,\mathbb{Z})\,,\\
\pi^*(G)& = \PD_{X_3}(\{\xi=0\}) \,\in\,H^2(X_3,\mathbb{Z})\,,
\end{align}
where $\xi$ is the homogeneous coordinate transverse to the O7, and $\pi$ is the orientifold projection. Let also $p$ be a $\mathbb{Z}_2$-parameter measuring the obstruction for $B_3$ to admit spin structure, namely:
\be\label{pBehavior}
p&=&\left\{\begin{array}{llll}0&&& w_2(B_3)=0\\1&&&w_2(B_3)\neq0\,.\end{array}\right.
\ee
Let finally $D$ be the Chern class of the normal bundle of the D7-branes stack inside $B_3$: 
\be
D&=&\PD_{B_3}(S_2)\,\in\,H^2(B_3,\mathbb{Z})\,,
\ee
where $S_2$ is the divisor of $X_3$ wrapped by the stack. In what follows, all the cohomology classes defined so far will be implicitly meant pulled-back to the Calabi-Yau threefold.

It is easy to find that the Sen weak coupling limit of the F-theory configuration with $Sp(N)$ singularity ($N\ge0$) on the divisor $S_2\subset X_3$ can be obtained upon tachyon condensation of a system of $2N+2$ D9-branes and $2N+2$ anti-D9-branes, with gauge bundles as follows:
\be \label{generalwhitneysum}
D9&&\quad\mathcal{O}\left(\frac{1}{2}(7-p)\,G-N\,D \right)\oplus\mathcal{O}\left(\frac{1}{2}(1-p)\,G \right)\oplus\0\\{}&& \quad\bigoplus_{i=1}^N\left[\mathcal{O}\left(i\,D-\frac{1}{2}(7+p)\,G \right)\oplus\mathcal{O}\left(\frac{1}{2}(7-p)\,G-(i-1)\,D\right)\right]\,,\0\\{}\0\\
\overline{D9}&&\quad\mathcal{O}\left(N\,D-\frac{1}{2}(7+p)\,G\right)\oplus\mathcal{O}\left(-\frac{1}{2}(1+p)\,G \right)\oplus\0\\{}&& \quad\bigoplus_{i=1}^N\left[\mathcal{O}\left(\frac{1}{2}(7-p)\,G-i\,D \right)\oplus\mathcal{O}\left((i-1)\,D-\frac{1}{2}(7+p)\,G\right)\right]\,,
\ee
and with a B-field of the form:
\be\label{campoB}
B=\frac{p}{2}\,G\,.
\ee
One can easily show that the above D9-$\overline{\rm D9}$ system induces the right lower D-brane charges, provided one computes them taking into account the B-field \eqref{campoB}, as in subsection \ref{necessaryB}. The D9-brane and D5-brane charges vanish; the D7-brane charge density is $8G$ distributed among the Whitney and the $Sp(N)$-stack; finally the total D3-brane charge density is independent of the values of $p,N$ and it reads:
\be\label{caricatotaleD3brana}
Q_{D3}&=&\frac{29}{2}\,G^3+\frac{1}{2}\,c_2(X_3)\,G\,.
\ee
This agrees with \eqref{finale} and \eqref{contributograv} for the smooth case, assuming that the Whitney umbrella has generic shape (saturation of the bound). The total D-brane charge (as seen by the double cover) is obtained by integrating \eqref{caricatotaleD3brana} over the Calabi-Yau threefold $X_3$.\\
The induced gauge flux on the stack of D7 and anti-D7-branes is the following:
\be\label{flussogaugegen}
F=\frac{1}{2}\,\sum_{i=1}^N\,((7-p)\,G-(2i-1)\,D)\,C_{2i-1}\;\quad,\;\quad (C_{2i-1})_{jk}=\delta_{ij}\,\delta_{ik}-\delta_{i+N,j}\,\delta_{i+N,k}\;,
\ee
where $G$ and $D$ are meant restricted to $S_2$. Again this flux breaks the gauge group $Sp(N)$ down to $SU(N)\times U(1)$. Note that the quantization condition of the gauge flux \eqref{flussogaugegen} is still regulated only by the even/odd-ness of the class $D$ of the stack, i.e. by the first Chern class of the normal bundle of $S_2$ in $X_3$. Indeed, the first term in the expression \eqref{flussogaugegen} is always an even class, due to \eqref{pBehavior}. Hence, the reduction modulo $2$ of the cohomology class $\left((7-p)G-(2i-1)D\right)|_{S_2}$ is $w_2\left(N_{X_3}S_2\right)=w_2(S_2)$. Therefore, the gauge flux  \eqref{flussogaugegen} is the right one to cancel the Freed-Witten anomaly of the D7 stack wrapping $S_2$, having holonomy in the class $w_2(S_2)$ \cite{Freed:1999vc,Bonora:2008hm}.

The gauge contribution to the D3-brane charge density induced by the flux \eqref{flussogaugegen} is:
\be\label{caricagaugeD3brana}
Q_{gauge}&=&\frac{N}{4}\,D\,\left[(7\,G-N\,D)^2+\frac{N^2-1}{3}\,D^2\right]\,.
\ee
Then, clearly, the gravitational contribution reads: 
\be
Q_{grav}&=&Q_{D3}-Q_{gauge}\,,
\ee
with $Q_{D3}$ as in \eqref{caricatotaleD3brana}. Putting $N=1$ in the square brackets of eq. \eqref{caricagaugeD3brana}, one recognizes the same structure of $(\Delta c_2)^2$, with $\Delta c_2$ as in Fulton's general formula \eqref{formulagenerale}. It would be interesting to find an iteration of Fulton's formula in the case of more than one blow-up ($N>1$) and verify that its structure agrees with eq. \eqref{caricagaugeD3brana}.

\section{Summary of results and outlook} \label{summaryoutlook}
In this section, we summarize our main results for the convenience of the reader, and conclude with an outlook on our forthcoming work.

\begin{itemize}

\item For any smooth, elliptically fibered Calabi-Yau fourfold $Z_4$ \emph{with a globally defined Weierstrass representation}, we find that
\begin{equation*}
\frac{c_2(Z_4)}{2} \in H^4(Z_4, \mathbb{Z})\,.
\end{equation*}

Since choosing $G_4=c_2(Z_4)/2$ would correspond to a IIB flux that breaks 4-dimensional Poincar\'e invariance, this result implies that such fluxes, being always integrally quantized, can be shifted to zero. Hence, no smooth F-theory fourfolds are ruled out.

This result also implies that no shifted quantization rule for the 7-brane gauge flux arises when the F-theory fourfold is free of non-abelian singularities.

\item A bulk B-field must be switched on in the type IIB weak coupling limit of an F-theory compactification, whenever the orientifold O7-plane has an odd $c_1(O7)$.

\item For an $Sp(N)$ (I$_{2N}^{ns}$) Kodaira singularity on $P_n=0\subset\PP^3$, with $\PP^3$ the base of the elliptic fourfold, the total induced physical D3-brane charge (i.e. as computed on $\PP^3$) is:
\be
n_{D3}&=& \frac{\chi(\tilde Z_4)}{24} -\tfrac{1}{2}\,\int_{Z_4} G_{\rm min} \wedge G_{\rm min} = 972\,,\0
\ee
where $\tilde Z_4$ is the resolved fourfold, and $G_{\rm min} = \Delta c_2/2$ is the minimal, obligatory, half-integrally quantized flux arising from the resolution of the singular $Z_4$.

The flux-induced tadpole is:
\be
n_{D3}^{\textrm{flux}}&=&\frac{nN}{4}\left[(28-nN)^2+n^2\,\frac{N^2-1}{3}\right]\,,\0
\ee
while the remainder is due to gravitational interactions.

The gauge flux on the D7-brane stack generating the above contribution to the D3 tadpole is:
\be
F=\frac{1}{2}H\,\sum_{i=1}^N\,(28-n(2i-1))\,C_{2i-1}\qquad\textrm{with}\qquad (C_{2i-1})_{jk}=\delta_{ij}\,\delta_{ik}-\delta_{i+N,j}\,\delta_{i+N,k}\;,\0
\ee
$H$ being the hyperplane class of the `upstairs' CY threefold $W\PP^3_{1,1,1,1,4}[8]$, and $C_{2i-1}$ the $2N\times2N$ Cartan matrices of  $Sp(N)$. Such a flux breaks the gauge group $Sp(N)$ to $SU(N)\times U(1)$. Results for the smooth configuration are recovered by just putting $N=0$.

\item For an $Sp(N)$ singularity on a divisor $S_2$ of a general K\"ahler base manifold $B_3$, and a double cover CY threefold $X_3$, the induced physical D3-brane charge (i.e. as measured on $B_3$) is:
\be
n_{D3}&=&\frac{1}{4}\,\int_{X_3}29\,\pi^*c_1(B_3)^3+c_2(X_3)\,\pi^*c_1(B_3)\,,\0
\ee
where $\pi:X_3\rightarrow B_3$ is the double cover projection. The gauge contribution to this D3 tadpole is:
\be
n_{D3}^{gauge}&=&\frac{N}{8}\,\int_{X_3}\pi^*\left\{\PD_{B_3}S_2\,\left[\left(7\,c_1(B_3)-N\,\PD_{B_3}S_2\right)^2+\frac{N^2-1}{3}\,(\PD_{B_3}S_2)^2\right]\right\}\,,\0
\ee
where $\PD$ denotes Poincar\'e duality. The gravitational contribution is simply $n_{D3}^{grav}=n_{D3}-n_{D3}^{gauge}$.

Finally, the gauge flux on the D7-brane stack wrapping the divisor $S_2$ is:
\be
F=\frac{1}{2}\,\sum_{i=1}^N\,\left[(7-p)\,c_1(B_3)-(2i-1)\,\PD_{B_3}S_2\right]|_{S_2}\,C_{2i-1}\;\;\;,\;\; (C_{2i-1})_{jk}=\delta_{ij}\,\delta_{ik}-\delta_{i+N,j}\,\delta_{i+N,k}\;.\0
\ee
Again such a flux breaks the gauge group $Sp(N)$ down to $SU(N)\times U(1)$. The bulk $B$-field reads:
\be
B&=&\frac{p}{2}\,c_1(B_3)\,,\0
\ee
where $p=0,1$ according to whether $B_3$ is spin or not respectively. \\
Again results for the smooth configuration are recovered by just putting $N=0$.

\end{itemize}

\paragraph{}

We would like to stress that these results constitute a non-trivial check of the intuition behind the correspondence between M/F-theory and IIB fluxes. As stated in the introduction, the sharp distinction between IIB bulk fluxes and 7-brane worldvolume fluxes is blurred in F/M-theory. Moreover, some 7-branes fluxes can be delocalized altogether when uplifted to $G_4$-fluxes in F/M-theory, making it harder to track the physical origin of a shift in their quantization.

Clearly the first and most important thing to do is to generalize our results to singularities other than of $C_N$-type. This will shed light on the mechanism of Freed-Witten anomaly cancellation for general non-perturbative bound states of 7-branes with general gauge group.

So far we have only been addressing the second of the two impacts mentioned at the beginning of section \ref{settingproblem} of a ``generalized'' Freed-Witten anomaly for F-theory, namely, the effects on flux quantization rules. \\
Actually, the M-theory G-flux suffers from another subtle topological effect, specifically, the necessary and sufficient (torsion-like) condition for anomaly cancellation. Indeed, Witten has conjectured \cite{Witten:1999vg} that the topological type of the G-flux restricted to the worldvolume of an M5-brane is constrained by a torsion class in the fourth integral cohomology of the M5 worldvolume. It would be very interesting to investigate the consequences that such condition has in the F-theory context, including its effects on flux quantization conditions.

\subsection*{Acknowledgements}
We would like to thank Ralph Blumenhagen, Andreas Braun, Volker Braun, Ilka Brunner, I\~naki Garcia-Etxebarria, Thomas Grimm, Benjamin Jurke, Luca Martucci, Christoph Mayrhofer, Gary Shiu and Roberto Valandro for useful discussions and explanations. R. S. would like to thank Loriano Bonora and Fabio Ferrari Ruffino for useful insights and encouragement, the Arnold-Sommerfeld-Center for Theoretical Physics at the LMU-Munich for the kind hospitality during the early stage of this work and the International School for Advanced Studies (SISSA) of Trieste with the Consorzio Erasmus Job Placement ``Key to Europe" for financial support. The work of A. C. is supported by a EURYI award of the European Science Foundation, and in part by the Excellence Cluster Universe, Garching.

\appendix

\section{Toric resolutions}\label{toricblowups}

In this appendix we present some basic techniques of toric geometry for illustrative purposes. They should be sufficient to understand all the steps of the blow-up processes performed in section \ref{SingularCase} to resolve non-abelian singularities. For a more complete treatment, the reader is referred to \cite{FultonToric,Hori:2003ic}.

We will construct the fan of the ambient variety of $K3$, which can be easily visualized, as a $W\PP^2_{2,3,1}$-fibration over the 2-sphere. Then, we will force the Kodaira singularity of type I$_2$ on a toric divisor of the base and deduce the projective weight assignments for the blown-up $K3$. Finally, and most importantly, we will show that the resolved $K3$ is still an elliptic fibration over the 2-sphere, but with a \emph{two}-component fiber: The affine node and the Cartan node of the extended Dynkin diagram of $SU(2)$. 

Consider the following parameterization of the $K3$ surface:
\be\label{weightsK3app}
\begin{array}{ccccc|ccc}x_1&x_2&X&Y&Z&&\textrm{Weierstrass}&{}\\ \hline
1&1&0&0&-2&&0\\
0&0&2&3&1&&6&\end{array}\begin{array}{r}\\ \\ ,\end{array}
\ee
where the homogeneous coordinates of the base $\PP^1$ are called $x_1,x_2$. The Weierstrass polynomial describing $K3$ as a divisor of the ambient threefold, will generically be a sum of monomials like the following
\be\label{generalmonomial}
x_1^{a+1}\,X^{b+1}\,Y^{c+1}\,Z^{d+1}\,x_2^{e+1}\,,
\ee 
with $a,b,c,d,e$ integer numbers. First of all, these numbers are not all independent, due to the two projective rescalings. Thus, compatibility with \eqref{weightsK3app} clearly requires:
\be
\left\{\begin{array}{rrr}a+e-2d&=&0\\ 2b+3c+d&=&0\end{array}\right.\qquad\Longrightarrow\qquad\left\{\begin{array}{lll}d&=&-2b-3c\\ e&=&-a-4b-6c\end{array}\right.\,.
\ee 
Moreover, they are subjected to several conditions, which can be represented as vectors in an integral three dimensional\footnote{The dimension of the lattice is always equal to the dimension of the ambient space.} lattice in the following way. Write the generic condition as:
\be\label{coordinatesfan}
w_1a+w_2b+w_3c\ge-1\,.
\ee
Then, the lattice vector corresponding to a given condition will have coordinates $(w_1, w_2, w_3)$.

There are conditions which are always present. They guarantee that the Weierstrass equation is well-defined, namely, the positivity of every exponent in \eqref{generalmonomial}. Thus, one has:
\be\label{K3fan}
\begin{array}{lllllclrlll}a\ge-1&&&(1&,&0&,&0)&&&x_1\\ b\ge-1&&&(0&,&1&,&0)&&&X\\ c\ge-1&&&(0&,&0&,&1)&&&Y\\ d\ge-1&&&(0&,&-2&,&-3)&&&Z\\ e\ge-1&&&(-1&,&-4&,&-6)&&&x_2\;, \end{array}
\ee
where the third column assigns to each vector the coordinate whose exponent is displayed in the first column. For a smooth $K3$, this is the end of the story, and the vectors in \eqref{K3fan} make up the so called \emph{fan} of the ambient toric variety of the $K3$. 

For each further condition imposed, one gets a new vector with a new associated coordinate. The new vector represents some constraint on the coefficients of the Weierstrass polynomial, which may generate a singularity of the $K3$. The new vector together with the previous ones, make up the fan of the blown-up ambient threefold.

From the coordinates of the vectors in \eqref{K3fan}, one realizes that the vector associated to $Z$ comes between the ones associated to $x_1$ and $x_2$, thus breaking the convex cone made by the last two. This implies that $x_1 x_2$ is an element of the Stanley-Reisner ideal of this toric variety, and thus these two coordinates make up a 2-sphere. Moreover, it can be shown that the whole toric threefold is a fibration on such a $\PP^1$. This can be quickly seen by verifying that the projection along the line generated by $x_1$ does not destroy any cone. The fan of the fiber is finally singled out as made by the vectors in the kernel of such a projection. These are obviously $X$, $Y$ and $Z$ (they all have zero in the first entry), and consequently $XYZ$ constitutes the other element of the Stanley-Reisner ideal.

Now it is the moment to force the type I$_2$ singularity for example on the toric divisor $x_1=0$ of the base. In analogy to \eqref{weiestrass}, the singularity of $K3$ will be located on the codimension three place in the ambient threefold (i.e. a point) described by $X=Y=x_1=0$. According to table \ref{TateClassif}, the only new condition to impose is that each monomial in the Weierstrass equation should be \emph{at least quadratic} in those three coordinates. Therefore, looking at \eqref{generalmonomial} and \eqref{coordinatesfan}, the right lattice vector to be added is $(1,1,1)$ and the associated coordinate will be called $v$. One is blowing up the locus $X=Y=x_1=0$ in the original ambient threefold because $v$ destroys the cone given by those three coordinates and the element $XYx_1$ must be added to the Stanley-Reisner ideal. Therefore, knowing all the lattice vectors, one can easily construct the table of projective weights for the blown-up ambient threefold and for the proper transform equation, which will no longer have the Weierstrass representation:
\be\label{weightsK3bupapp}
\begin{array}{cccccc|ccc}x_1&x_2&X&Y&Z&v&&\textrm{proper transform}&{}\\ \hline
1&1&0&0&-2&0&&0\\
0&0&2&3&1&0&&6&\\
1&0&1&1&0&-1&&2\end{array}\begin{array}{r}\\ \\ \\.\end{array}
\ee
Again, this toric manifold is a fibration over a 2-sphere, whose coordinates now are $(vx_1,x_2)$, as they both have projective weights $(1,0,0)$ in \eqref{weightsK3bupapp}. One can deduce that  the Stanley-Reisner ideal gains two elements, $vx_2$ $vZ$.

The fan of the fiber is still made by the vectors corresponding to $X$, $Y$ and $Z$. The fiber of the blown-up $K3$ over a generic point of the base is everywhere a $T^2$, given by the proper transform on that point, except for the locus $vx_1=0$. Indeed, as it is manifest, on this base point the fiber splits into two parts given by the proper transform in which one puts $v=0$ and $x_1=0$ respectively. The $x_1=0$ component has the topology of a degree \emph{one} $\PP^1$. The $v=0$ component, instead, is a \emph{quadratic} $\PP^1$, in the $\PP^2$ defined by the coordinates $x_1,X,Y$. The two components intersect in two points, as seen by imposing  $v=x_1=0$ in the proper transform equation. The $v=0$ component is the true exceptional divisor and thus represents the Cartan node of the $SU(2)$ Dynkin diagram. The $x_1=0$ component represents the affine node of the extended Dynkin diagram of $SU(2)$.

As last useful information, the relation between this toric blow-up and the one performed with ``traditional'' methods \cite{GriffithsHarris} is as follows. Introduce a $\PP^2$ with homogeneous coordinates $s$, $t$ and $u$. This is the auxiliary space used to perform the traditional blow-up of codimension three locus. Then, the map which links the toric method to the latter is the following:
\be
\varphi\,:\,(x_1,x_2,X,Y,Z,v)\,\longrightarrow\,(\tilde{x}_1,\tilde{x}_2,\tilde{X},\tilde{Y},\tilde{Z},s,t,u)\,=\,(vx_1,x_2,vX,vY,Z,X,Y,x_1)\,.
\ee
Indeed, the table of projective weights of the new homogeneous coordinates is:
\be
\begin{array}{cccccccc}\tilde{x}_1&\tilde{x}_2&\tilde{X}&\tilde{Y}&\tilde{Z}&s&t&u\\ \hline
1&1&0&0&-2&0&0&1\\
0&0&2&3&1&2&3&0\\
0&0&0&0&0&1&1&1\end{array}\begin{array}{r}\\ \\ \\,\end{array}
\ee
which are exactly the right weight assignments for the homogeneous coordinates of a toric threefold blown-up with traditional techniques. They are compatible with the (two independent) blow-up constraints that impose $(s,t,u) \sim (X,Y,x_1)$, namely
\be
\textrm{rank}\,\left(\begin{array}{cc}s&X\\t&Y\\u&x_1\end{array}\right)&=&1\,.
\ee

\section{The type I$_2$ Kodaira singularity}\label{KodSU2}

The aim of this appendix is to describe in detail the geometrical features of the type I$_2$ Kodaira singularity and of its toric resolution in a simple toy model. For  notations and techniques adopted, see the appendix \ref{toricblowups}.

Tate's classification, prescribes to drop from the Weierstrass equation in the Tate form\footnote{See \cite{Bershadsky:1996nh} for the notations.}
\be \label{weiestrass}
Y^2 +a_1 XYZ + a_3 YZ^3= X^3 + a_2 X^2Z^2 +a_4 XZ^4 +a_6Z^6\,,
\ee 
every monomial that is not at least quadratic in those three coordinates. The analysis here is parallel to that in appendix \ref{toricblowups}, except for the two spectator base coordinates, $x_3$ and $x_4$. The fan of the resolved ambient toric fivefold $\tilde{M}_5$, which is a $W\PP^2_{2,3,1}$-bundle over $\PP^3$, is the following:
\be\label{Mtildefan}
\begin{array}{cc}\textrm{canonical}&\left\{\begin{array}{llccccclrlll}(1&,&0&,&0&,&0&,&0)&&&x_1\\ (0&,&1&,&0&,&0&,&0)&&&X\\ (0&,&0&,&1&,&0&,&0)&&&Y\\ (0&,&0&,&0&,&1&,&0)&&&x_3\\ (0&,&0&,&0&,&0&,&1)&&&x_4\\ (0&,&-2&,&-3&,&0&,&0)&&&Z\\ (-1&,&-8&,&-12&,&-1&,&-1)&&&x_2 \end{array}\right.\\ \\ \hline\\ \textrm{additional}&\begin{array}{llllccccclrllllll}&&&(1&,&1&,&1&,&0&,&0)&&&&& v \end{array} \end{array}\begin{array}{c}\\ \\ \\ \\ \\ \\ \\ \\ \\.\end{array}
\ee
The last two columns in \eqref{Mtildefan} play no essential role in the blow-up procedure, so one can disregard the coordinates, $x_3$ and $x_4$, which span the 4-5 plane in the five dimensional lattice. The variety represented by this fan is also readily recognized as a fibration on a $\PP^3$, with the following projection map:
\be
\pi\,:\,(x_1,x_2,x_3,x_4,X,Y,Z,v)\,\longrightarrow\,(\tilde{x}_1,\tilde{x}_2,\tilde{x}_3,\tilde{x}_4)\,=\,(vx_1,x_2,x_3,x_4)\,.
\ee
The fan of the fiber is generated by $X$, $Y$ and $Z$. From \eqref{Mtildefan} we deduce the table of projective weights for the ambient $\tilde{M}_5$ and for the proper transform which describes the blown-up Calabi-Yau fourfold as an hypersurface of $\tilde{M}_5$:
\be
\begin{array}{cccccccc|ccc}x_1&x_2&x_3&x_4&X&Y&Z&v&&\textrm{proper transform}&{}\\ \hline
1&1&1&1&0&0&-4&0&&0\\
0&0&0&0&2&3&1&0&&6\\
1&0&0&0&1&1&0&-1&&2\end{array}\begin{array}{r}\\ \\ \\.\end{array}
\ee
Finally, the Stanley-Reisner ideal is:
\be\label{StanleyReisnersu2}
\textrm{SR ideal :}\,\left\{x_1x_2x_3x_4\,;\,XYZ\,;\,x_1XY\,;\,x_2x_3x_4v\,;\,vZ\right\}\,.
\ee 

The equation of the proper transform is
\be\label{proptranssu2}
&Y^2+a_1(x_1v,x_i)XYZ+a_{3,1}(x_1v,x_i)x_1YZ^3&=vX^3+a_2(x_1v,x_i)X^2Z^2+\0\\&+a_{4,1}(x_1v,x_i)x_1XZ^4+a_{6,2}(x_1v,x_i)x_1^2Z^6&\qquad\qquad\qquad i=2,3,4\,,
\ee
This has manifestly lost the Weierstrass representation, where the polynomials depend on the coordinates of the new $\PP^3$ base.

As explained in appendix \ref{toricblowups}, the fiber is everywhere elliptic on the base except where the former singularity was, namely, in the new coordinates, $x_1v=0$. Here, the fiber splits in two components: $v=0$, is the fiber of the exceptional divisor of the blow-up, and it represents the Cartan node of the $SU(2)$ Dynkin diagram. The other component, $x_1=0$, is the affine node of the extended diagram, present also in the non-singular case, and it is fibered over the divisor $S_2\simeq\PP^2$ with coordinates $x_2$, $x_3$ and $x_4$. The exceptional divisor itself is also fibered over $S_2$, and it is given by simply substituting $v=0$ in eq. \eqref{proptranssu2}:
\be\label{cartannode}
Y^2+a_1(x_i)XYZ+a_{3,1}(x_i)x_1YZ^3=a_2(x_i)X^2Z^2+a_{4,1}(x_i)x_1XZ^4+a_{6,2}(x_i)x_1^2Z^6\,,
\ee
where $i=2,3,4$. Since $Z$ must be different from zero when $v=0$ (see \eqref{StanleyReisnersu2}), one can gauge fix $Z=1$ in the above equation, which, afterwards, appears as an irreducible non-singular quadratic equation describing a $\PP^1$ of degree \emph{two} in the $\PP^2$ of coordinates $(x_1,X,Y)$. The other component, namely the affine curve fibered over $S_2$ is obtained by substituting $x_1=0$ in eq. \eqref{proptranssu2}:
\be\label{affinenode}
Y^2+a_1(x_i)XYZ=vX^3+a_2(x_i)X^2Z^2\,,
\ee
where again $i=2,3,4$. Notice that $X$ must be different from zero in the above equation, otherwise $Y$ would be forced to be also vanishing, which is not allowed by  \eqref{StanleyReisnersu2}. Therefore, one can gauge fix $X=1$ and realize that the above equation will become linear and describe a degree \emph{one} $\PP^1$ with coordinates $(Y,Z)$ by eliminating the coordinate $v$ in favor of them.\\

It is worth to probe a bit more this illustrative example, in order to see the properties of the intersections of these two components and describe their relation with D7-branes.

Here, we are really dealing with the type I$_2^{ns}$ Kodaira singularity (see table \ref{TateClassif}, taken from \cite{Bershadsky:1996nh}), although in the general treatment of geometric singularities the separation between split and non-split case starts at the next step of Tate's algorithm, that is I$_3$ for this branch. Indeed, as it appears from the table, already for I$_2$ the two cases differ for the order of zero of $a_2$: If $a_2=0\;\textrm{mod}\,x_1$, then one has the true $SU(2)$ gauge group, otherwise, in case $a_2$ is completely generic, one has the group $Sp(1)$ which is anyhow isomorphic to $SU(2)$. The I$_2^{ns}$ has a perturbative realization, whereas the I$_2^{s}$ does not (see for instance \cite{Donagi:2009ra} for $SU(5)$).\\
\begin{table}[ht]
\centering
\begin{tabular}{|c|c|c|c|c|c|c|c|}
\hline
 type & group & $ a_1$ &$a_2$ & $a_3$ &$ a_4 $& $ a_6$ &$\Delta$ \\ \hline \hline I$_0 $ & --- &$ 0 $ &$ 0
$ &$ 0 $ &$ 0 $ &$ 0$ &$0$ \\ \hline I$_1 $ & --- &$0 $ &$ 0 $ &$ 1 $ &$ 1
$ &$ 1 $ &$1$ \\ \hline I$_2 $ &$SU(2)$ &$ 0 $ &$ 0 $ &$ 1 $ &$ 1 $ &$2$ &$
2 $ \\ \hline I$_{3}^{ns} $ & unconven. &$0$ &$0$ &$2$ &$2$ &$3$ &$3$ \\ \hline
I$_{3}^{s}$ &$SU(3)$ &$0$ &$1$ &$1$ &$2$ &$3$ &$3$ \\ \hline
I$_{2k}^{ns}$ &$ Sp(k)$ &$0$ &$0$ &$k$ &$k$ &$2k$ &$2k$ \\ \hline
I$_{2k}^{s}$ &$SU(2k)$ &$0$ &$1$ &$k$ &$k$ &$2k$ &$2k$ \\ \hline
I$_{2k+1}^{ns}$ &unconven. & $0$ &$0$ &$k+1$ &$k+1$ &$2k+1$ &$2k+1$ \\ \hline
I$_{2k+1}^s$ &$SU(2k+1)$ &$0$ &$1$ &$k$ &$k+1$ &$2k+1$ &$2k+1$ \\ \hline
II & --- &$1$ &$1$ &$1$ &$1$ &$1$ &$2$ \\ \hline
III &$SU(2)$ &$1$&$1$ &$1$ &$1$ &$2$ &$3$ \\ \hline
IV$^{ns} $ &unconven. &$1$ &$1$ &$1$&$2$ &$2$ &$4$ \\ \hline
IV$^{s}$ &$SU(3)$ &$1$ &$1$ &$1$ &$2$ &$3$ &$4$\\ \hline
I$_0^{*\,ns} $ &$G_2$ &$1$ &$1$ &$2$ &$2$ &$3$ &$6$ \\ \hline
I$_{2k-3}^{*\,ns}$ &$SO(4k+1)$ &$1$ &$1$ &$k$ &$k+1$&$2k$ &$2k+3$ \\  \hline
I$_{2k-3}^{*\,s}$ &$SO(4k+2)$ &$1$ &$1$ &$k$ &$k+1$
&$2k+1$ &$2k+3$ \\ \hline I$_{2k-2}^{*\,ns}$ &$SO(4k+3)$ &$1$ &$1$ &$k+1$
&$k+1$ &$2k+1$ &$2k+4$ \\ \hline I$_{2k-2}^{*\,s}$ &$SO(4k+4)^*$ &$1$ &$1$
&$k+1$ &$k+1$ &$2k+1$ 
&$2k+4$ \\ \hline IV$^{*\,ns}$ &$F_4 $ &$1$ &$2$ &$2$ &$3$ &$4$
&$8$\\ \hline IV$^{*\,s} $ &$E_6$ &$1$ &$2$ &$2$ &$3$ &$5$ & $8$\\ \hline
III$^{*} $ &$E_7$ &$1$ &$2$ &$3$ &$3$ &$5$ & $9$\\  \hline II$^{*} $
&$E_8\,$ &$1$ &$2$ &$3$ &$4$ &$5$ & $10$ \\ \hline
 non-min & --- &$ 1$ &$2$ &$3$ &$4$ &$6$ &$12$\\ \hline
\end{tabular}\vspace{0.2cm}\caption{\small Tate's algorithm. When the group is not specified it is Abelian, and the $*$ on the groups $SO(4k+4)$ means that a further factorization condition must be imposed.}\label{TateClassif}
\end{table}
Nevertheless, the difference between the two cases shows up upon analyzing the intersection between the Cartan node and the affine node of the blown-up fiber. Such an intersection is a codimension three locus in $\tilde{M}_5$ (made by two points fibered over $S_2$, as shown later) and, in the non-split case, it reads:
\be\label{intersectionspalle}
\left\{\begin{array}{l}
v=0\\ x_1=0\\ Y^2+a_1(x_i)XYZ-a_2(x_i)X^2Z^2=0\quad\qquad i=2,3,4\,.\end{array}\right.
\ee
The difference between split and non-split resides in the possibility of factorizing the above polynomial: In this more generic non-split case, such polynomial cannot be factorized, and monodromies occur along the brane worldvolume $S_2$. Notice, however, that here what are undergoing monodromies are actually the two intersection points between the two components of the blown-up fiber, rather than the two components themselves (as in the following non-split cases of the algorithm). So there is no folding of the Dynkin diagram, the latter being made only of one Cartan node (${\bf A_1}$). 

On the other hand, in the less generic split case, the last monomial in \eqref{intersectionspalle} becomes $a_{2,1}(x_i)x_1vX^2Z^2$, and thus the polynomial factorizes. Therefore, no monodromy will exchange the two intersection points.

To see more geometrically this codimension three manifold described by \eqref{intersectionspalle}, first go to a basis in which the blown-up ambient fivefold and the proper transform have the following weights:
\be
\begin{array}{cccccccc|ccc}x_1&x_2&x_3&x_4&X&Y&Z&v&&\textrm{proper transform}&{}\\ \hline
1&1&1&1&8&12&0&0&&24\\
0&0&0&0&2&3&1&0&&6\\
1&0&0&0&1&1&0&-1&&2\end{array}\begin{array}{r}\\ \\ \\.\end{array}
\ee
Since, by \eqref{StanleyReisnersu2}, $Z\neq0$ in \eqref{intersectionspalle}, one can fix the second gauge by taking $Z=1$. Then, taking into account $v=x_1=0$, the residual SR-ideal will be made by $XY$ and $x_2x_3x_4$. Thus, the ambient toric threefold is a $\PP^1_{X,Y}$-bundle over $S_2\simeq\PP^2_{x_2,x_3,x_4}$, as can be seen below:
\be\label{ambientthreefoldinters}
\begin{array}{ccccc|c}x_2&x_3&x_4&X&Y&\textrm{proper transform}\\ \hline
1&1&1&8&12&24\\
0&0&0&1&1&2\end{array}&\sim&\begin{array}{ccccc|c}x_2&x_3&x_4&X&Y&\textrm{proper transform}\\ \hline
1&1&1&0&4&8\\
0&0&0&1&1&2\end{array}\begin{array}{r}\\ \\ \\.\end{array}
\ee
On this ambient threefold one is imposing what remains of the proper transform, namely the equation $Y^2+a_1XY-a_2X^2=0$. Here, one can gauge fix the last $\mathbb{C}^*$-action (second row in \eqref{ambientthreefoldinters}) by taking $X=1$. Finally, one arrives at the following degree eight equation in the previous ambient threefold, seen now as $W\PP^3_{1114}$:
\be\label{Z2fibration}
Y^2+a_1(x_i)Y-a_2(x_i)=0\quad\qquad i=2,3,4\,.
\ee
We can see that the intersection between the exceptional divisor and the affine one is a $\mathbb{Z}_2$-fibration over $S_2$, where, over each point of the brane worldvolume the fiber consists in $Y_\pm$, which are the solutions of \eqref{Z2fibration}. The fiber degenerates in one point on the curve given by the vanishing locus of the discriminant of \eqref{Z2fibration}, i.e. $a_1^2+4a_2=0$. 

What is the physics behind all that? These two points are locally described by the two different functions $Y_\pm(x_i)$, and represent the positions of the two D7-branes making together the stack on which the $SU(2)$ gauge theory lives. On the curve $a_1^2(x_i)+4a_2(x_i)=0$, the two points degenerate into one, $Y_+(x_i)=Y_-(x_i)=-a_1(x_i)/2$.

In Sen's weak coupling limit, the two D7's in question are geometrically invariant under the orientifold involution. However, the orientifold action swaps them. Therefore, the O7 resides where the D7-brane intersects its image, namely where $Y_+$ and $Y_-$ coincide. Hence, the O7 will wrap the degree eight divisor $h\equiv a_1^2+4a_2$ inside $B_3$. 
In type IIB, however, the two D7-branes coincide on the non-singular divisor $x_1=0$. In the F-theory lift, they are only separated along the fiber direction. 

A priori, one could expect that the enhanced singularities along higher codimension loci on the $SU(2)$ 7-brane are not resolved by just blowing-up the $SU(2)$-singularity as we described. In appendix \ref{compresol} we prove that they \emph{do} get resolved, and no further singularity remains.

\section{Smoothness of the blown-up fourfold}\label{compresol}

In this appendix, we present a simple calculation to prove that, in the case of I$_2^{ns}$ singularities, the blow-up induced by the lattice vector $v=(1,1,1,0,0)$ is sufficient to \emph{completely} resolve the elliptic Calabi-Yau fourfold. In other words, by solving the singularity over the $SU(2)$ 7-brane, all other singularity enhancements at higher codimension are automatically resolved too.

It is expected that the same conclusion holds for worse singularities, provided all the blow-up's induced by the relevant lattice vectors are performed. A list of such vectors corresponding to any given Kodaira singularity can be found in \cite{Bershadsky:1996nh}.

Suppose we have an $SU(2)$ singularity on $x_1=0\subset\PP^3$. Consider the proper transform of the elliptic fibration after the blow-up induced by $v$:
\be\label{trasfpropria}
&Y^2+a_1(x_1v,x_i)XYZ+a_{3,1}(x_1v,x_i)x_1YZ^3&=vX^3+a_2(x_1v,x_i)X^2Z^2+\0\\&+a_{4,1}(x_1v,x_i)x_1XZ^4+a_{6,2}(x_1v,x_i)x_1^2Z^6&\qquad\qquad\qquad i=2,3,4\,,
\ee
and compute its gradient $\vec{\nabla}$. Since any possible residual singularity in \eqref{trasfpropria} must lie on the exceptional divisor, it suffices to restrict the gradient to the submanifold $v=0$. Gauge-fixing $Z=1$, one obtains:
\be
\vec{\nabla}|_E=\left(\begin{array}{c}a_1Y-2a_2X-a_{4,1}x_1\\ 2Y+a_1X+a_{3,1}x_1\\ a_1XY+3a_{3,1}x_1Y-2a_2X^2-4a_{4,1}x_1X-6a_{6,2}x_1^2\\ a_{3,1}Y-a_{4,1}X-2a_{6,2}x_1\\ \d_ia_1XY+\d_ia_{3,1}x_1Y-\d_ia_2X^2-\d_ia_{4,1}x_1X-\d_ia_{6,2}x_1^2\\ \d_1a_1x_1XY+\d_1a_{3,1}x_1^2Y-X^3-\d_1a_2x_1X^2-\d_1a_{4,1}x_1^2X-\d_1a_{6,2}x_1^3 \end{array}\right)\begin{array}{c}\\ \\ \\ \\ \\ , \end{array}
\ee
where the eight rows are the derivatives of \eqref{trasfpropria} with respect to $X,Y,Z,x_1,x_i,v$ respectively and $\d_1$ means the derivative of the polynomials with respect to the first argument.
In analogy with the notations of \cite{Bershadsky:1996nh}, define:
\be
b_2&\equiv&a_1^2+4a_2\,,\0\\
b_{4,1}&\equiv&a_1a_{3,1}+2a_{4,1}\,,\0\\
b_{6,2}&\equiv&a_{3,1}^2+4a_{6,2}\,,\0\\
\ee
where $b_2=h$ is the O7-plane in the weak coupling limit. Then, after some algebraic manipulations, the conditions of vanishing of the gradient read:
\be\label{condizionising}\left\{
\begin{array}{rcl}Y&=&-\frac{a_1X+a_{3,1}x_1}{2}\\ b_2X+b_{4,1}x_1&=&0\\ b_{4,1}X+b_{6,2}x_1&=&0\\ b_2X^2+4b_{4,1}x_1X+3b_{6,2}x_1^2&=&0\\ 4X^3+(\d_1b_2X^2+2\d_1b_{4,1}x_1X+\d_1b_{6,2}x_1^2)x_1&=&0\\ \d_ib_2X^2+2\d_ib_{4,1}x_1X+\d_ib_{6,2}x_1^2&=&0\,.
\end{array}\right.\ee
From the above equations, we see that residual singularities of the blown-up fourfold (if any) cannot lie on the intersection between the exceptional divisor and the affine component of the resolved fiber, $x_1=0$. Indeed, the fourth equation would imply $X=0$ and so, by the first equation, $Y$ should vanish too, which is impossible in the ambient variety.\\
Therefore, one can gauge-fix in \eqref{condizionising} $x_1=1$. Now a case-by-case analysis is needed.
\begin{itemize}
\item If $X=0$, then $b_{4,1}=b_{6,2}=\d_Ib_{6,2}=0$ ($I=1,i$), which are too many conditions to be imposed on $S_2$, and thus generically no solution is found.
\item If $X\neq0$, but $b_2=0$, then again there are too many conditions on $S_2$, namely $b_2=b_{4,1}=b_{4,1}=0$, plus others coming from the derivatives.
\item  If $X\neq0$ and also $b_2\neq0$, then $X=-b_{4,1}/b_2$, with $b_{4,1}\neq0$. Both the third and the fourth equation above implie $b_{8,2}\equiv b_2b_{6,2}-b_{4,1}^2=0$, which restricts to a curve in $S_2$. But then four further equations on the derivatives have to be imposed, which generically do not intersect. 
\end{itemize}

This concludes the proof. No residual singularities survive after this toric blow-up of the original elliptic, I$_2^{ns}$-singular Calabi-Yau fourfold.


\providecommand{\href}[2]{#2}\begingroup\raggedright\endgroup

\end{document}